\documentclass[aps,pra,reprint,amsmath,amssymb,superscriptaddress,onecolumn,longbibliography,notitlepage,floatfix]{revtex4-2}
\usepackage[nearskip,margin = 0pt,caption=false]{subfig}
\usepackage[hidelinks]{hyperref}
\usepackage{physics}
\usepackage{tikz}
\usepackage[labelfont=bf, justification=raggedright, figurename={Fig. }, tablename={Table}, labelsep=period, font=small]{caption}

\begin{document}
\title{Complete vectorial optical mode converter using multi-layer metasurface}
\author{Go~Soma}
\email{go.soma@tlab.t.u-tokyo.ac.jp}
\affiliation{School of Engineering, The University of Tokyo, 7-3-1 Bunkyo-ku, Tokyo 113-8656, Japan}
\author{Kento~Komatsu}
\affiliation{School of Engineering, The University of Tokyo, 7-3-1 Bunkyo-ku, Tokyo 113-8656, Japan}
\author{Yoshiaki~Nakano}
\affiliation{School of Engineering, The University of Tokyo, 7-3-1 Bunkyo-ku, Tokyo 113-8656, Japan}
\author{Takuo~Tanemura}
\email{tanemura@ee.t.u-tokyo.ac.jp}
\affiliation{School of Engineering, The University of Tokyo, 7-3-1 Bunkyo-ku, Tokyo 113-8656, Japan}

\begin{abstract} 
A vectorial optical mode converter that can transform an orthogonal set of multiple input vector beams into another orthogonal set of vector beams is attractive for a wide range of applications in optics and photonics.
While multi-plane light conversion (MPLC) and metasurface (MS) technologies have been explored to individually address multiple spatial mode conversion and polarization mode manipulation, there has been no universal methodology to simultaneously convert a set of multiple vectorial modes, having non-uniform spatial distributions in both their complex amplitude and polarization, to another set of multiple vectorial modes. 
In this paper, we demonstrate versatile devices based on the MPLC concept incorporating multi-layer locally birefringent MSs and present a general design formalism for complete vectorial mode conversion in arbitrary cases. 
The effectiveness of our proposed method is confirmed experimentally by demonstrating a 6-mode (3 spatial modes $\times$ 2 polarization modes) multiplexer, fabricated on a compact chip with a $\sim$0.65~mm$^2$ lateral size in a folded MS configuration.
Additionally, we verify its applicability to more advanced functional devices through numerical demonstration of a mode-division-multiplexed dual-polarization coherent receiver and spatial-mode-multiplexed vectorial holography. 
%Moreover, the final device was fabricated on a compact single chip and experimentally demonstrated. 
The versatility of our protocol makes it suitable for designing a myriad of multi-input-multi-output devices, providing a powerful tool for realizing universal optical mode converters for a wide range of applications.
\end{abstract}
\maketitle
%%%%%%%%%%%%%%%%%%%%%%%%%%  body  %%%%%%%%%%%%%%%%%%%%%%%%%%
Every linear optical component can be considered as a mode converter that transforms a specific set of input orthogonal modes into another set of output orthogonal modes \cite{Miller2012-hl}. 
The optical modes generally provide the fundamental basis for describing complex optical systems in an explicit and economical manner \cite{Miller2019-AOP}. To enjoy the maximum degrees of freedom (DoFs) of light in a given optical system, therefore, full usage of optical modes, including the spatial and polarization modes in addition to the wavelength, is essential. In space-division multiplexed (SDM) optical communication, for example, a large number of dual-polarization spatial modes are multiplexed in an optical fiber to increase the transmission capacity \cite{Mizuno2016-na, Winzer2017-tx, Soma2018-el, Rademacher2021-bn, Sillard2022-vi, van-den-Hout2024-zz}.
The use of multiple vectorial optical modes with non-uniform spatial profiles in both their complex amplitude and polarization, such as cylindrical vector beams (CVBs) \cite{Zhan2009-gd}, has also been explored to boost the information density of free-space optical (FSO) communication \cite{Milione2015-yt, Zhu2021-gy, He2020-wc, Chen2021-sa} and holographic imaging \cite{Song2022-wb, Xiong2023-mk, Dorrah2021-yb, Ren2020-qm, Bao2022-mz} systems.
It is, therefore, of paramount interest in optics and photonics to develop a universal converter that can transform an arbitrary set of orthogonal vectorial modes with generally non-uniform polarization profiles into another set of orthogonal vectorial modes.
In particular, an ideally lossless device that achieves unitary multi-input-multi-output (MIMO) vectorial mode conversion is attractive for diverse applications.

To this end, multi-plane light conversion (MPLC) is a promising approach that has been successful in achieving universal spatial mode conversion for scalar optical fields \cite{Morizur2010-zn, Labroille2014-vy, Zhang2023-lu}.
Through a succession of transverse phase manipulation and free-space propagation, an arbitrary orthogonal set of spatial modes can be converted into another set of spatial modes in a unitary manner \cite{Morizur2010-zn}.
Using a spatial light modulator (SLM) and a multi-reflecting mirror, highly scalable mode (de)multiplexers for hundreds of optical modes have been demonstrated \cite{Fontaine2019-pw, Fontaine2021-im}.
Such MPLC concept has also been demonstrated on integrated photonic platforms, where various types of multi-input optical mode mixers were employed to replace free-space propagation \cite{Tang2017-PTL, Tang2018-OL, Tang2021-mh, Tanomura2022-PRL, Tanomura2023-jz}.
Owing to the inherent scalability and excellent performances, MPLC devices have widely been used for various applications, including optical communication \cite{Rademacher2021-bn, Labroille2014-vy, Tanomura2023-jz, Zhang2020-zd, Wen2019-gk, Wen2020-ih, Lengle2016-po, Fang2021-lk, Zhang2023-lu}, %, Oh2022-tp, Oh2024-ux}, 
quantum optics \cite{Brandt2020-kp}, and optical computing \cite{Tang2021-mh, Lin2018-mm}. 

Despite these advantages, however, conventional MPLC schemes have been limited to the conversion of scalar optical fields.
%Indeed, polarization-independent phase plates or SLMs are commonly used as phase masks, which cannot manipulate the vectorial fields arbitrarily and convert them into different vector beams.
%Therefore, to achieve dual-polarization operation, the polarization-diversity scheme needs to be employed, where the input beam is split into two polarization states and independent spatial mode conversion is applied to each of them using an MPLC device \cite{Mounaix2020-cx, Li2023-yt}. Such an approach, however, suffers from additional complexity and bulkiness. Moreover
While a polarization-diversity scheme can be employed to apply independent spatial mode conversion to two polarization states \cite{Mounaix2020-cx, Li2023-yt}, such an approach cannot transform non-uniform spatial distributions of polarization in an arbitrary manner, which is insufficient to achieve complete conversion of multiple vector beams in general cases.

On the other hand, all-dielectric metasurfaces (MSs) have been studied actively over the past decade as efficient flat optical devices that can manipulate the polarization properties of incident beams \cite{Yu2014-we, Khorasaninejad2017-sb, Arbabi2022-jk}. They are composed of two-dimensional (2D) arrays of sub-wavelength scatterers called meta-atoms, each of which generally has asymmetric geometries and functions as an ultra-small birefringent material. 
%Each meta-atom with asymmetric geometry functions as an ultra-small birefringent material and can provide various Jones-matrix operations to the local field of light \cite{Balthasar_Mueller2017-ex, Arbabi2015-rv}. 
Through judicious design of meta-atoms, therefore,
a variety of polarization-dependent properties can be obtained, such as polarization beam splitting and polarization-dependent holographic imaging \cite{Rubin2019-tb, Arbabi2018-vz, Fan2020-tq, Bao2021-yc, Bao2022-mz, Zheng2022-cb, Soma2023-fx}.
%While such mono-layer (or bi-layer) MS devices usually assume an input beam with a single spatial mode, it was also demonstrated that simultaneous manipulation of multiple spatial/polarization modes can be achieved for some specific sets of orthogonal modes, such as optical angular momentum (OAM) modes and Fourier modes that have spatial symmetries \cite{Jin2019-tl, Ren2019-cw, Ren2020-bs, Zhou2020-vl, Kamali2017-jt, Jang2021-ss, Deng2023-mj}.
%Such mono-layer (or bi-layer) MS devices, however, basically assume an input beam with a single spatial mode, and is not capable of converting multiple spatial and polarization modes unless for some specific cases, such as optical angular momentum (OAM) modes and Fourier modes that have spatial symmetries \cite{Jin2019-tl, Ren2019-cw, Ren2020-bs, Zhou2020-vl, Kamali2017-jt, Jang2021-ss, Deng2023-mj}.
Whereas these previous demonstrations using mono-layer or bi-layer MS devices employed an input beam with a single spatial mode, simultaneous conversion of multiple vectorial modes with arbitrary spatial and polarization profiles 
%is generally not possible using a mono-layer MS and 
requires beam propagation through cascaded layers of MSs.
However, versatile multi-layered MS devices to achieve universal and simultaneous MIMO vectorial mode conversions for arbitrary cases have not been demonstrated to our knowledge.

In this paper, we present a fully vectorial mode converter using a multi-layer MS and provide a general design formalism to realize desired MIMO vectorial mode conversions for arbitrary cases.
We combine the concepts of MPLC and MS by replacing the scalar phase masks in conventional MPLC devices with locally birefringent MSs. 
%Due to the high spatial resolution and versatile functionalities, the presented device can achieve all functionalities, such as deflection, focusing, collimation, and polarization rotation, without additional free-space optics. 
As a result, the conventional MPLC theory is extended to include multiple stages of Jones matrices.
We then derive an explicit inverse design protocol based on the adjoint method to optimize all the meta-atoms in each MS layer so that a target mode conversion is obtained.
%The presented scheme is validated by demonstrating three unique devices. 
The presented concept is validated experimentally by demonstrating a 6-mode (3 spatial modes $\times$ 2 polarization modes) multiplexer using a 4-layer MS, fabricated on a compact $\sim$0.65~mm$^2$ chip with a folded MS configuration.
%is fabricated in a folded MS configuration to successfully demonstrate simultaneous spatial/polarization mode conversion to $X$ and $Y$ linearly polarized (LP) modes for all six inputs.
Furthermore, the applicability of our scheme to more advanced functional devices is verified numerically by demonstrating a mode-division-multiplexed (MDM) dual-polarization coherent receiver and spatial-mode-multiplexed vectorial holography with excellent performances.
%First, a mode-division-multiplexed (MDM) dual-polarization coherent receiver is designed to detect six coherent signals (3 spatial modes $\times$ 2 polarization modes). By optimizing six layers of MSs, all six signal modes, as well as the local oscillator (LO) mode, are focused onto 24 well-defined spots at the output with precise intensities, phases, and polarizations. As a result, simultaneous balanced homodyne detection of all signals is obtained with outstanding performances, such as 0.9-dB insertion losses and 0.1-dB mode-dependent losses (MDLs).
%Outstanding performances, such as 0.9-dB insertion losses and 0.1-dB mode-dependent losses (MDLs), are obtained numerically for all modes, which are comparable or superior to those of commercial single-channel coherent receivers.
%Second, unique spatial-mode-multiplexed vectorial holography is realized by a 4-layer MS, where eight different holographic images are generated depending on the input CVB mode and the analyzing polarization state at the output. Such a device is not possible with a mono-layer MS and has never been demonstrated to the best of our knowledge.
%Finally, spatial (LP$_{01}$, LP$_{\rm 11a}$, and LP$_{\rm 11b}$) and polarization ($X/Y$) mode multiplexer is designed and fabricated in the folded MS configuration with a four-layer MS. By using the fabricated compact chip with a lateral size below 0.6 mm$^2$, desired LP and polarization modes were experimentally observed for all $X$-polarized mode inputs from a single-mode fiber (SMF).
Owing to the versatility of the presented formalism, a variety of vectorial mode converters can be realized to utilize the full DoFs of optical beams for diverse applications.

\section*{Multi-input vectorial mode converter using a multi-layer metasurface}
\begin{figure}[t!]
\centering\includegraphics{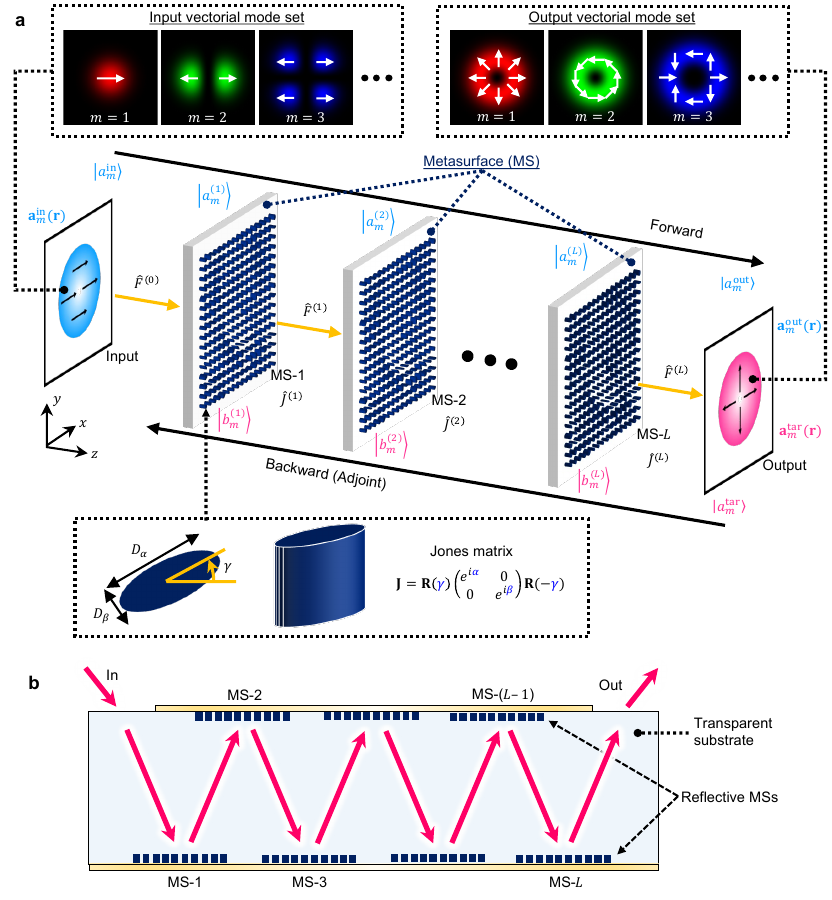}
\caption{
\textbf{Universal vectorial mode conversion using a multi-layer MS.} 
{\bf a}, Schematic of the vectorial mode converter using an $L$-layer metasurface (MS). The input vectorial mode set $\qty{\ket{a_m^{\mathrm{in}}}|m=1,2,...,M}$ is transformed into another mode set $\qty{\ket{a_m^{\mathrm{out}}}|m=1,2,...,M}$ by repeated free-space propagation $\hat{F}^{(l)}$ and lightwave conversion by MSs $\hat{J}^{(l)}$. The Jones matrices of MSs are optimized using the forward and adjoint (backward) fields. The bottom inset shows schematics of the birefringent meta-atom. Arbitrary phase shifts $\alpha$ and $\beta$ can be obtained by properly designing meta-atom dimensions, $D_\alpha$ and $D_\beta$. 
{\bf b}, Schematic of a single-chip mode converter with a folded MS configuration using reflective MSs.}
\label{concept}
\end{figure}
Figure~\ref{concept}a shows the schematic of the vectorial mode converter using $L$ layers of MSs.
By repeating multiple stages of lightwave conversion through the birefringent MS and propagation through the free space, we can achieve simultaneous conversions of $M$ orthogonal input vectorial modes to $M$ desired output vectorial modes, including their polarization profiles.
In practice, this device can be implemented by a single compact chip with the folded MS configuration \cite{Faraji-Dana2018-xc, Faraji-Dana2019-ox, Oh2022-tp} as shown in Fig.~\ref{concept}b. 

The input vectorial field of the $m$-th mode ($m=1,2,\dots,M$) can be written in a Jones-vector form as
\begin{equation}
    \vb{a}_{m}^{\mathrm{in}}(\vb{r})=\mqty(a_{m,X}^{\mathrm{in}} (\vb{r})\\a_{m,Y}^{\mathrm{in}} (\vb{r})),
    \label{eq:j_in}
\end{equation}
where $\vb{r}=(x,y)^t$ denotes the in-plane position and $a_{m,X/Y}^{\mathrm{in}}(\vb{r})$ represents the $X/Y$-polarized components of the field. 
(Note that we employ uppercase $X$ and $Y$ to represent the polarization and lowercase $x$ and $y$ to represent the in-plane position in this paper to avoid confusion.)
Here, we assume that the in-plane variation of the field is gradual compared with the wavelength so that the $z$ component of the electromagnetic field can be ignored.
%(i.e., paraxial approximation).
For convenience, we use the Dirac notation to represent the input field as
\begin{equation}
    \ket{a_m^{\mathrm{in}}}
    =\sum_{n,p} a_{m,p}^{\mathrm{in}}(\vb{r}_n)\ket{n,p}.
    \label{eq:a_in}
\end{equation}
Note that the in-plane position $\vb{r}$ is discretized into $N$ points, $\vb{r}_n=(x_n,y_n)^t$ ($n=1,2,\dots, N$), and $\ket{n,p}$ ($n=1,2,\dots, N;\ p=X, Y$) represents the orthonormal basis in terms of position and polarization. 
For convenience, $\vb{r}_n$ are matched to the center positions of periodically placed meta-atoms.
Following successive free-space propagation and light conversion by the MSs, the resulting vectorial field at the input of the $l$-th MS can be expressed as 
\begin{equation}
%    \ket{a_m^{(l)}}
%    =\sum_{n,p} a_{m,p}^{(l)}(\vb{r}_n)\ket{n,p}
%    =\hat{F}^{l-1} \hat{J}^{l-1} \cdots \hat{J}^1 \hat{F}^{0}\ket{a_m^{\mathrm{in}}},
    \ket{a_m^{(l)}}
    =\sum_{n,p} a_{m,p}^{(l)}(\vb{r}_n)\ket{n,p}
    =\hat{F}^{(l-1)} \hat{J}^{(l-1)} \cdots \hat{F}^{(1)}\hat{J}^{(1)} \hat{F}^{(0)}\ket{a_m^{\mathrm{in}}},
    \label{eq:a_ml}
\end{equation}
where $a_{m,p}^{(l)}(\vb{r}_n)\equiv \langle n, p | a_m^{(l)}\rangle$ denotes the $p$-polarized field at $\vb{r}_n$ for the $m$-th input mode. The operator $\hat{F}^{(l)}$ describes free-space propagation from the output of the $l$-th MS to the input of the $(l+1)$-th MS and is written as 
\begin{equation}
    \hat{F}^{(l)}
    =\sum_{n,n',p}f_{nn'}^{(l)}\ketbra{n,p}{n',p}.
    \label{eq:F_hat}
\end{equation}
Here, $f_{nn'}^{(l)}\equiv\langle{n,p}|\hat{F}^{(l)}|{n',p}\rangle$ physically represents the coupling coefficient from $\vb{r}_{n'}$ at the $l$-th plane to $\vb{r}_n$ at the $(l+1)$-th plane, which can be described by the Rayleigh–Sommerfeld point-spread function (impulse response).
Since $X$ and $Y$ polarization components of light independently follow the Helmholtz equations as they propagate through a uniform medium, they do not couple during free-space propagation. We thus have $\bra{n,p}\hat{F}^{(l)}\ket{n',p'}=0\ (p\neq p')$.
Finally, the operator $\hat{J}^{(l)}$ in Eq.~(\ref{eq:a_ml}) represents the propagation through the $l$-th MS and can be written as
\begin{equation}
    \hat{J}^{(l)}
    =\sum_{n,p,p'}j_{pp'}^{(l)}(\vb{r}_n)\ketbra{n,p}{n,p'},
    % =\mqty(
    % \vb{J}_{1}^{(l)} & & O \\
    %  & \ddots &  \\
    % O &  & \vb{J}_{N}^{(l)}
    % ),
    \label{eq:J_hat}
\end{equation}
where 
% $j_{pp'}^{(l)}(\vb{r}_n)\equiv\mel**{n,p}{\hat{J}^{(l)}}{n,p'}$
$j_{pp'}^{(l)}(\vb{r}_n)\equiv\langle{n,p}|{\hat{J}^{(l)}}|{n,p'}\rangle$ denotes the conversion of the Jones vector induced by the meta-atom at $\vb{r}_n$. Since each meta-atom can only locally transform the Jones vector at its position, $j_{pp'}^{(l)}(\vb{r}_n)\equiv\langle{n,p}|{\hat{J}^{(l)}}|{n',p'}\rangle\ (n\neq n')$
% $\mel**{n,p}{\hat{J}^{(l)}}{n',p'}=0\ (n\neq n')$. 
%Once again, we assume that the in-plane variation of $j_{pp'}^{(l)}(\vb{r}_n)$ is gradual compared with the wavelength so that we can ignore the $Z$-polarized component after transmitting through the MS.
For convenience, we define the Jones matrix induced by each meta-atom at $\vb{r}_n$ on the $l$-th MS as
\begin{equation}
    \vb{J}^{(l)}(\vb{r}_n)
    =\mqty(
    j_{XX}^{(l)}(\vb{r}_n) & j_{XY}^{(l)}(\vb{r}_n) \\
    j_{YX}^{(l)}(\vb{r}_n) & j_{YY}^{(l)}(\vb{r}_n)
    ).
    \label{eq:J_matrix}
\end{equation}
Assuming an ideally lossless and non-chiral dielectric structure as shown in the bottom inset of Fig.~\ref{concept}a, each meta-atom functions as an ultra-small birefringent waveplate. Thus, $\vb{J}^{(l)}(\vb{r}_n)$ can be written explicitly as \cite{Balthasar_Mueller2017-ex}%:
\begin{equation}
    \vb{J}^{(l)}(\vb{r}_n) = \vb{R}\mqty(\gamma^{(l)}(\vb{r}_n))
    \mqty(e^{i\alpha^{(l)}(\vb{r}_n)} & 0 \\ 0 & e^{i\beta^{(l)}(\vb{r}_n)})
    \vb{R}\mqty(-\gamma^{(l)}(\vb{r}_n)),
    \label{eq:J_atom}
\end{equation}
where $\vb{R}(\gamma)$ is a rotation matrix defined as
\begin{equation}
    \vb{R}(\gamma) \equiv \mqty(\cos{\gamma} & -\sin{\gamma} \\ \sin{\gamma} & \cos{\gamma}).
    \label{eq:R}
\end{equation}
In Eq.~(\ref{eq:J_atom}), $\alpha^{(l)}(\vb{r}_n)$ and $\beta^{(l)}(\vb{r}_n)$ represent the phase shifts for the eigenmode waves polarized along the slow and fast axes of the meta-atom, respectively, and $\gamma^{(l)}(\vb{r}_n)$ is the angle of orientation. Arbitrary phase shifts $\alpha$ and $\beta$ can be obtained by judiciously selecting the dimensions of the meta-atom $(D_\alpha, D_\beta)$ along the two axes as shown in the bottom inset of Fig.~\ref{concept}a \cite{Arbabi2015-rv, Soma2023-fx}.
Therefore, each meta-atom can be described using three design parameters: $\alpha$, $\beta$, and $\gamma$.

\section*{Adjoint optimization of metasurface}
We now consider an efficient algorithm to optimize the parameters of each meta-atom so that an objective function $\mathcal{E}$ is maximized.
Here, $\mathcal{E}$ is defined by the averaged inner product between the output vectorial field $\ket{a_m^{\mathrm{out}}}$ and the target field $\ket{a_m^{\mathrm{tar}}}$ for all $M$ modes as
\begin{equation}
    \mathcal{E}\equiv \frac{1}{M} \sum_m \qty|\bra{a_m^{\mathrm{tar}}}\ket{a_m^{\mathrm{out}}}|^2
    =\frac{1}{M} \sum_{m} \qty|\sum_{n,p}\qty(a_{m,p}^{\mathrm{tar}}(\vb{r}_n))^*a_{m,p}^{\mathrm{out}}(\vb{r}_n)|^2,
    \label{eq:objective}
\end{equation}
where 
\begin{align}
    \ket{a_m^{\mathrm{out}}} 
    &= \sum_{n,p} a_{m,p}^{\mathrm{out}}(\vb{r}_n)\ket{n,p}
    =\hat{F}^{(L)}\hat{J}^{(L)}\ket{a_m^{(L)}}, \\
    \ket{a_m^{\mathrm{tar}}}
    &= \sum_{n,p} a_{m,p}^{\mathrm{tar}}(\vb{r}_n)\ket{n,p}.
\end{align}
%* is the complex conjugate. 
Note that $a_{m,p}^{\mathrm{out/tar}}(\vb{r}_n)\equiv\langle{n,p}|{a_{m}^{\mathrm{out/tar}}}\rangle$ represents the output/target $p$-polarized component of the $m$-th mode. 

To maximize $\mathcal{E}$, we employ the adjoint method \cite{Molesky2018-uw, Oh2022-tp}; the design parameters of each meta-atom are updated iteratively as
\begin{equation}
    \theta^{(l)}(\vb{r}_n) \leftarrow \theta^{(l)}(\vb{r}_n)+\mathcal{F}\qty[\pdv{\mathcal{E}}{\theta^{(l)}(\vb{r}_n)}],
    \label{eq:p_opt}
\end{equation}
where $\theta^{(l)}(\vb{r}_n)\in\qty{\alpha^{(l)}(\vb{r}_n),\beta^{(l)}(\vb{r}_n),\gamma^{(l)}(\vb{r}_n)}$ ($n=1,2,\dots,N$) represents the parameters of the meta-atom at $\vb{r}_n$ in the $l$-th MS and
$\mathcal{F}$ is an optimization function of the first-order gradient, which is defined appropriately to achieve rapid convergence.

After some mathematical procedures (See Supplementary Note~1 for the complete derivation), we can derive
\begin{equation}
\begin{split}
    \pdv{\mathcal{E}}{\theta^{(l)}(\vb{r}_n)}=\frac{2}{M}\sum_{m} \Re \qty[
    \braket{a_m^{\mathrm{out}}}{a_m^{\mathrm{tar}}}
    \mel**{b_m^{(l)}}{\pdv{\hat{J}^{(l)}}{\theta^{(l)}(\vb{r}_n)}}{a_m^{(l)}}
    ].
    % &=\frac{1}{M}\sum_{m} 2\Re \qty[
    % \braket{a_m^{\mathrm{out}}}{a_m^{\mathrm{tar}}}
    % \sum_{p,p'}
    % \pdv{j_{pp'}^{(l)}(\vb{r}_n)}{\theta^{(l)}(\vb{r}_n)}
    % \braket{b_m^{(l)}}{n,p}
    % \braket{n,p'}{a_m^{(l)}}
    % ]\\
    % &=\frac{1}{M}\sum_{m} 2\Re \qty[
    % \braket{a_m^{\mathrm{out}}}{a_m^{\mathrm{tar}}}
    % \sum_{p,p'}
    % \pdv{j_{pp'}^{(l)}(\vb{r}_n)}{\theta^{(l)}(\vb{r}_n)}
    % \qty(b_{m,p}^{(l)}(\vb{r}_n))^*
    % a_{m,p'}^{(l)}(\vb{r}_n)
    % ]\\
%    &=\frac{2}{M}\sum_{m} \Re \qty[
%    \braket{a_m^{\mathrm{out}}}{a_m^{\mathrm{tar}}}
%    \{\vb{b}_{m}^{(l)}(\vb{r}_n)\}^\dagger \pdv{\vb{J}^{(l)}(\vb{r}_n)}{\theta^{(l)}(\vb{r}_n)} \vb{a}_{m}^{(l)}(\vb{r}_n)
%    ].
\end{split}
    \label{eq:dEdp3}
\end{equation}
%Note that we employ Eq.~(\ref{eq:J_matrix}) and define $\vb{a}_{m}^{(l)}(\vb{r}_n)$ and $\vb{b}_{m}^{(l)}(\vb{r}_n)$ as
%\begin{equation}
%    \vb{a}_{m}^{(l)}(\vb{r}_n) \equiv \mqty(a_{m,X}^{(l)}(\vb{r}_n)\\a_{m,Y}^{(l)}(\vb{r}_n)), \
%    \vb{b}_{m}^{(l)}(\vb{r}_n) \equiv \mqty(b_{m,X}^{(l)}(\vb{r}_n)\\b_{m,Y}^{(l)}(\vb{r}_n)).    
%\end{equation}
Here, $\langle b_m^{(l)}|$
%$\bra{b_m^{(l)}}$
is the adjoint vectorial field and  
%$|b_m^{(l)}\rangle$ is 
defined as
\begin{equation}
    \ket{b_m^{(l)}}\equiv \sum_n b_{m,p}^{(l)}(\vb{r}_n)\ket{n,p} 
    \equiv \hat{F}^{(l)\dagger}\cdots\hat{J}^{(L)\dagger}\hat{F}^{(L)\dagger}\ket{a_m^{\mathrm{tar}}},
    \label{eq:adjoint}
\end{equation}
which represents the vectorial field on the output of the $l$-th MS when the target field $\ket{a_m^{\mathrm{tar}}}$ is propagated backward as shown in Fig.~\ref{concept}a. 

Using Eq.~(\ref{eq:J_atom}), $\pdv{\vb{J}^{(l)}}{\theta^{(l)}}$ in Eq. (\ref{eq:dEdp3}) can be expressed explicitly (see Supplementary Note~1 for the actual expressions).
%From Eq.~(\ref{eq:J_atom}), $\pdv{\vb{J}^{(l)}}{\theta^{(l)}}$ can be expressed explicitly as
%\begin{align}
%    \pdv{\vb{J}^{(l)}(\vb{r}_n)}{\alpha^{(l)}(\vb{r}_n)}&=ie^{i\alpha^{(l)}(\vb{r}_n)}\mqty(\cos^2\gamma^{(l)}(\vb{r}_n) & \frac{1}{2}\sin 2\gamma^{(l)}(\vb{r}_n) \\ \frac{1}{2}\sin 2\gamma^{(l)}(\vb{r}_n) & \sin^2\gamma^{(l)}(\vb{r}_n)),
%    \label{eq:dJdp1}\\
%    \pdv{\vb{J}^{(l)}(\vb{r}_n)}{\beta^{(l)}(\vb{r}_n)}&=ie^{i\beta^{(l)}(\vb{r}_n)}\mqty(\sin^2\gamma^{(l)}(\vb{r}_n) & \frac{1}{2}\sin 2\gamma^{(l)}(\vb{r}_n) \\ \frac{1}{2} \sin 2\gamma^{(l)}(\vb{r}_n) & \cos^2\gamma^{(l)}(\vb{r}_n)),
%    \label{eq:dJdp2}\\
%    \pdv{\vb{J}^{(l)}(\vb{r}_n)}{\gamma^{(l)}(\vb{r}_n)}&=\qty(e^{i\alpha^{(l)}(\vb{r}_n)}-e^{i\beta^{(l)}(\vb{r}_n)})\mqty(-\sin 2\gamma^{(l)}(\vb{r}_n) & \cos 2\gamma^{(l)}(\vb{r}_n) \\ \cos 2\gamma^{(l)}(\vb{r}_n) & \sin 2\gamma^{(l)}(\vb{r}_n)).
%    \label{eq:dJdp3}
%\end{align}
Hence, $\pdv{\mathcal{E}}{\theta^{(l)}}$
%the derivatives of $\mathcal{E}$ 
for all MS parameters ${\theta^{(l)}}$ can be obtained at once from Eq.~(\ref{eq:dEdp3}) by computing $|a_m^{(l)}\rangle$ and $|b_m^{(l)}\rangle$ ($m=1,2,\dots, M$) through the forward and backward propagation.
%The actual optimization is performed as follows (detailed explanations of the optimization procedure are given in Supplementary Fig.~1). 
%Initially, the MS parameters of $\alpha^{(l)}(\vb{r}_n)$, $\beta^{(l)}(\vb{r}_n)$ and $\gamma^{(l)}(\vb{r}_n)$ are uniformly set to 0, $\pi/2$, and $\pi/4$, respectively, to prevent zero gradients in this work.
In each iteration of optimization, we first calculate $|a_m^{(l)}\rangle$ for each mode by the forward propagation given by Eq.~(\ref{eq:a_ml}) and derive $\mathcal{E}$ using Eq.~(\ref{eq:objective}). 
Similarly, $|b_m^{(l)}\rangle$ are obtained by Eq.~(\ref{eq:adjoint}).
%Here, we use the angular spectrum method (ASM) \cite{Matsushima2009-ba, Matsushima2010-wd} in the calculation of the free-space propagation.
Then, $\pdv{\mathcal{E}}{\theta^{(l)}(\vb{r}_n)}$ are calculated using Eq.~(\ref{eq:dEdp3}).
%with Eq.~(\ref{eq:dJdp1})-(\ref{eq:dJdp3}). 
Finally, we update the parameters through Eq.~(\ref{eq:p_opt}).
%In this work, we use the adaptive moment estimation (ADAM) algorithm \cite{Kingma2014-hl} as an optimizer for efficient updating of parameters.
These procedures are repeated until the objective function converges (see Supplementary Fig.~1).

\section*{Experimental results}

\begin{figure}[t!]
\centering\includegraphics{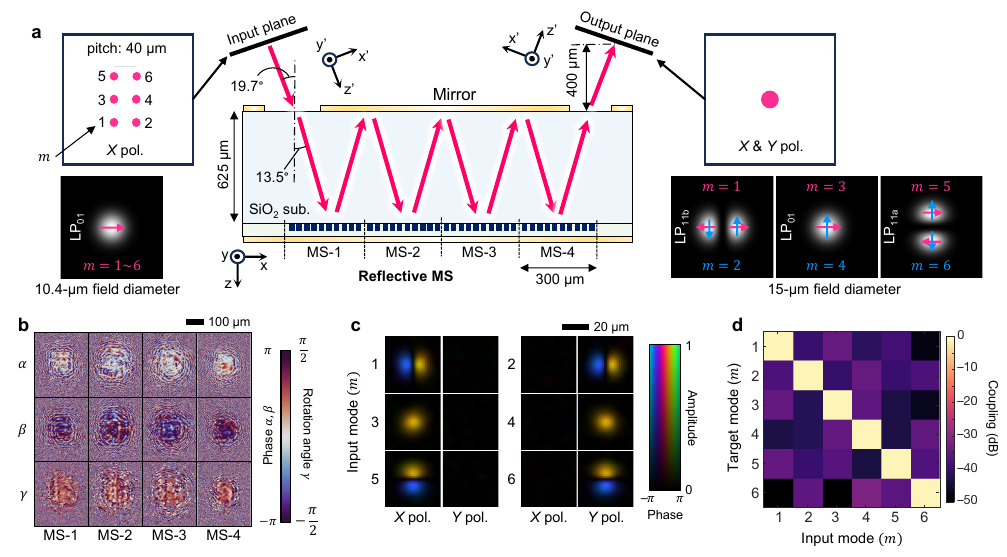}
\caption{
{\bf Schematic and design of a spatial/polarization mode multiplexer.} 
{\bf a}, Device configuration. Through multiple reflections at four reflective MS sections, six $X$-polarized input beams are converted to polarization-multiplexed LP modes at the output. 
{\bf b}, Spatial distributions of MS parameters $\alpha^{(l)}(\vb{r})$, $\beta^{(l)}(\vb{r})$, and $\gamma^{(l)}(\vb{r})$ ($l=1,2,3,4$) after optimization. %Scale bar is 100 {\textmu}m.
{\bf c}, Output complex field profiles for each input mode. %Scale bar is 10 {\textmu}m.
{\bf d}, Calculated coupling efficiency matrix $\vb{C}$.}
\label{MUX_sim}
\end{figure}

\begin{figure}[tb!]
\centering\includegraphics{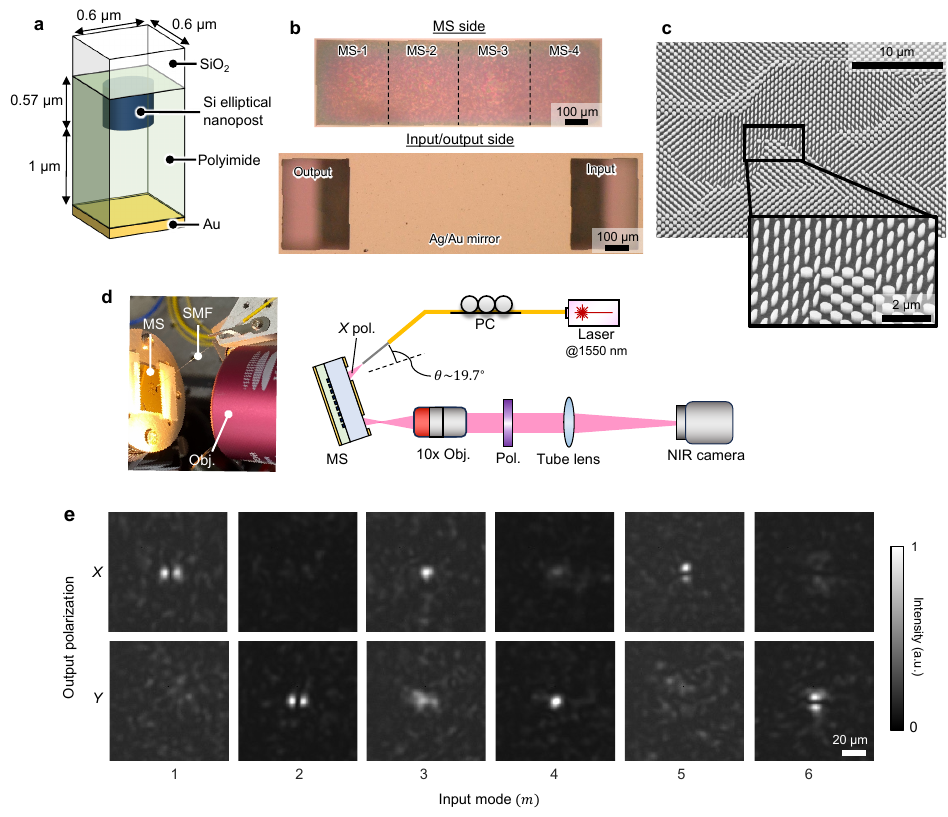}
\caption{
{\bf Experimental demonstration of a spatial/polarization mode multiplexer.} 
{\bf a}, Schematic of the reflective meta-atom, composed of a Si elliptical nanoposts on a SiO$_2$ substrate with polyimide and Au mirror layers. 
{\bf b}, Microscope images of the fabricated folded MS device at the MS side before Au mirror deposition (top) and at the input/output side (bottom). %Scale bars are 100 {\textmu}m. 
{\bf c}, Scanning electron microscope (SEM) images of the fabricated MS before depositing the polyimide layer. 
{\bf d}, Measurement setup. 
% The polarizer and reference path were used for the measurement of intensity and complex-field distributions, respectively. 
PC:~polarization controller. Obj.:~objective lens. Pol.:~polarizer. The left inset shows a photograph of the mode multiplexer chip with the input single-mode fiber (SMF) and the objective lens.
{\bf e}, Measured intensity profiles at the output plane in $X$ and $Y$ polarization. 
% {\bf f}, Measured coupling matrix $\vb{C}$.
}
\label{MUX_exp}
\end{figure}

To validate the generalized formalism and the optimization method presented in the previous section, we first consider a simple example of a 6-mode (3 spatial modes $\times$ 2 polarization modes) multiplexer.
%Here, we consider a device that simultaneously converts six $X$-polarized light beams at different input positions to three spatial modes (LP$_{01}$, LP$_{\rm 11a}$, and LP$_{\rm 11b}$) in two polarization ($X/Y$).
Figure~\ref{MUX_sim}a shows the schematic of the device. 
It is composed of a row of four reflective MS sections integrated on one side of a 625-{\textmu}m-thick fused silica (SiO$_2$) substrate and a total-reflection mirror layer with input/output apertures on the other side. 
Figure \ref{MUX_exp}a shows the reflective meta-atom employed in this work, which consists of a Si nanopost with a height of 0.57~{\textmu}m, capped with a polyimide layer and a gold mirror layer. 
The lateral dimension of each MS section is chosen to be 300~{\textmu}m $\times$ 360~{\textmu}m, which is sufficiently larger than forward-propagated beams at the first MS layer and backward-propagated beams at the last MS layer for all modes.
The incident light transmitted through the input aperture is reflected back and forth between the MS and mirror layers and exits through the output aperture.
We aim to design each MS section so that six $X$-polarized Gaussian beams ($m=1,...,6$) arranged on an equally spaced 2$\times$3 array at the input plane are converted into polarization-multiplexed spatial modes at the output plane.
As the output spatial modes, we assume LP modes inside a few-mode fiber (FMF) with a mode field diameter (MFD) of 15~{\textmu}m, supporting three spatial modes (LP$_\mathrm{01}$, LP$_\mathrm{11a}$, and LP$_\mathrm{11b}$) at 1550-nm wavelength.
As shown in the left and right insets of Fig.~\ref{MUX_sim}a, $X$-polarized input beams at $m=1$, $3$, and $5$ ($2$, $4$, and $6$) are converted to $X$-polarized ($Y$-polarized) LP$_{11b}$, LP$_{01}$, and LP$_{11a}$ modes, respectively, centered at the same position.
Owing to the folded MS configuration, simultaneous MIMO vectorial mode conversion can be achieved using a single compact chip.

%First, we numerically derived the spatial distributions of design parameters ($\alpha$, $\beta$, and $\gamma$) at four MS sections.
%% following the optimization procedure outlined in the previous section (see Methods for the details of simulation).
%The discretization spacing of the position $\vb{r}_n$ in the simulation was set to 1.2〜{\textmu}m (i.e., $N=250\times300$), which corresponded to twice the actual meta-atom spacing to be fabricated (0.6〜{\textmu}m) to save the optimization cost.
%Thus, the entire device contained 900,000 ($= 4\times250\times300\times3$) parameters. 
Figures~\ref{MUX_sim}b and \ref{MUX_sim}c respectively show the MS parameters of the optimized design and the simulated vectorial field distributions at the output plane for each input mode  (see Methods and Supplementary Fig.~2 for the design procedure). We can confirm that each input mode is transformed to the desired spatial and polarization mode.
For quantitative evaluation, we employ the coupling efficiency matrix $\vb{C}$, whose components are defined as $C_{mm'}=\qty|\bra{a_m^{\mathrm{tar}}}\ket{a_{m'}^{\mathrm{out}}}|^2$. 
The calculated $\vb{C}$ for the optimized MS design is shown in Fig.~\ref{MUX_sim}d.
The insertion loss and crosstalk are suppressed below 0.95~dB and $-30$~dB for all six modes.
%We then derived the actual geometries of each meta-atom required to achieve optimized MS parameters given in Fig.~\ref{MUX_sim}b. 
%The reflective meta-atom employed in this work is shown schematically in Fig.~\ref{MUX_exp}a. It consists of a Si nanopost with a height of 0.57 {\textmu}m, capped with a polyimide layer and a gold mirror layer. 
%The Si nanopost has an elliptical structure with dimensions $D_\alpha$ and $D_\beta$ as shown in Fig.~\ref{concept}a to induce birefringence.  The meta-atoms were placed on a square lattice with a lattice constant of 0.6 {\textmu}m, which was less than the wavelength to avoid higher-order diffraction. 
%The optimized MS parameter distributions depicted in Fig.~\ref{MUX_sim}b were first interpolated onto a square grid with a 0.6 {\textmu}m spacing. Then, the meta-atom dimensions ($D_\alpha$, $D_\beta$) and rotation angle at each position were determined through rigorous coupled-wave analysis (RCWA) (see the Methods for the detailed procedure and Supplementary Fig.~2). 
%
%The designed folded MS device was fabricated by electron-beam (EB) lithography and reactive-ion-etching (RIE) processes (see the Methods for the detailed fabrication process and Supplementary Fig.~3).

Figure~\ref{MUX_exp}b shows the microscope images of the fabricated device (see Methods and Supplementary Fig.~3 for the details of the fabrication process). 
% The lateral dimension of the entire device is 1.2 mm $\times$ 0.36 mm. 
%The input and output apertures with a size of 380 {\textmu}m $\times$ 280 {\textmu}m were aligned with the MS patterns formed on the opposite side.
Scanning electron microscope (SEM) images of the MS before capping the polyimide layer are shown in Fig.~\ref{MUX_exp}c. %We can see that a dense array of elliptical Si nanoposts was fabricated nicely as designed.
The lateral dimension of the entire device is 1.8~mm $\times$ 0.36~mm $\sim$ 0.65~mm$^2$. 

The fabricated device was characterized at 1550-nm wavelength using the setup shown in Fig.~\ref{MUX_exp}d (see Methods for the details). %An $X$-polarized beam from a single-mode fiber (SMF) at 1550-nm wavelength was incident to the MS device at an angle of 19.7$^\circ$. The input mode ($m$) was controlled by moving the SMF using a fiber stage to the corresponding position. The output plane of the device was imaged with a near-infrared (NIR) camera using a custom-built microscope. The output polarization was selected by a polarizer. 
Figure~\ref{MUX_exp}e shows the $X$- and $Y$-polarized intensity profiles, which were observed for each input fiber position.
We can confirm that each input mode was converted to the corresponding LP and polarization mode as designed.
The total transmission loss from the input SMF to the output light was measured to be 4.0~dB for all six inputs.
The coupling efficiency matrix was calculated using the complex field profiles measured by the off-axis holography (detailed explanation is given in Supplementary Note~2 and Supplementary Fig.~4).
% Then, we performed the off-axis holography to reconstruct the complex field profiles (refer to Supplementary Fig.~5). Then, we calculated the coupling matrix $\vb{C}'$ where $C'_{mm'}=\qty|\bra{a_m^{\mathrm{tar}}}\ket{a_{m'}^{\mathrm{out,meas}}}|^2$ ($\ket{a_{m'}^{\mathrm{out,meas}}}$: the measured output mode for the $m'$-th mode input). Figure~\ref{fig7}e shows the calculated coupling matrix.
The total coupling loss to the designed LP modes was in the range of 13.8 to 17.7~dB.
%which includes the $\sim$4 dB loss within the device and other losses from 
This relatively large modal mismatch may result from various reasons.
First, fabrication errors of MS, such as variations in Si nanopost geometries and the polyimide layer thickness, should have caused undesired scattering and non-perfect conversion of optical modes.
In addition, various issues in the experiment, such as the deviation of the incident beam from the ideal Gaussian beam assumed in the design, possible misalignment particularly in the incident angle, and undesired reflection at the input facet, should have increased the error from the ideal case.
There is, therefore, a large room for improvement through optimizing the device fabrication and measurement system.
%In addition, the interactions between adjacent meta-atoms, which were not considered in our design, may also have caused errors in resultant Jones matrices.
%Further improvement can thus be expected through improved fabrication processes and advanced algorithms in determining meta-atom geometries \cite{Molesky2018-uw}.

\section*{Applications to functional multi-modal devices}
%To demonstrate the effectiveness of the generalized formalism and the optimization method presented in the previous section, we consider two specific applications: (i) an MDM dual-polarization coherent receiver and (ii) spatial-mode-multiplexed vectorial holography.
To further investigate the efficacy of our scheme for more advanced applications, we design and numerically demonstrate fully vectorial MIMO devices for two other use cases: (i) MDM dual-polarization coherent receiver and (ii) spatial-mode-multiplexed vectorial holography.

\subsection*{MDM dual-polarization coherent receiver}
\label{sec: MDM-CohRx}
MDM technology using a multi-mode fiber (MMF) is promising for future optical communication systems to break the limit of transmission capacity through an SMF \cite{Mizuno2016-na, Winzer2017-tx}. In a polarization-multiplexed coherent MDM system, the receiver requires complex optical components to demultiplex all space/polarization modes and interfere each of them with the LO light through an optical hybrid before the detection by balanced photodiodes (PDs).
While we have recently demonstrated the use of a mono-layer MS to achieve simultaneous detection of spatially separated multiple coherent signals from a multi-core fiber \cite{Komatsu2024-pi}, the same approach cannot be applied to demodulate MDM signals from an MMF that overlap heavily in space.
Conventionally, therefore, MDM coherent receivers were implemented using three separate devices: a spatial mode demultiplexer, a polarization-beam splitter (PBS), and an optical hybrid for each mode \cite{Soma2018-el, Rademacher2021-bn, Sillard2022-vi, van-den-Hout2024-zz}.
Recently, Wen {\it et al.} have proposed a scalar MPLC-based device that combines a spatial mode demultiplexer and 90$^\circ$ optical hybrids for a single polarization \cite{Wen2019-gk, Wen2020-ih}.
However, a single device that can receive dual-polarization MDM coherent signals has never been demonstrated to our knowledge.

\begin{figure*}[t!]
\centering\includegraphics{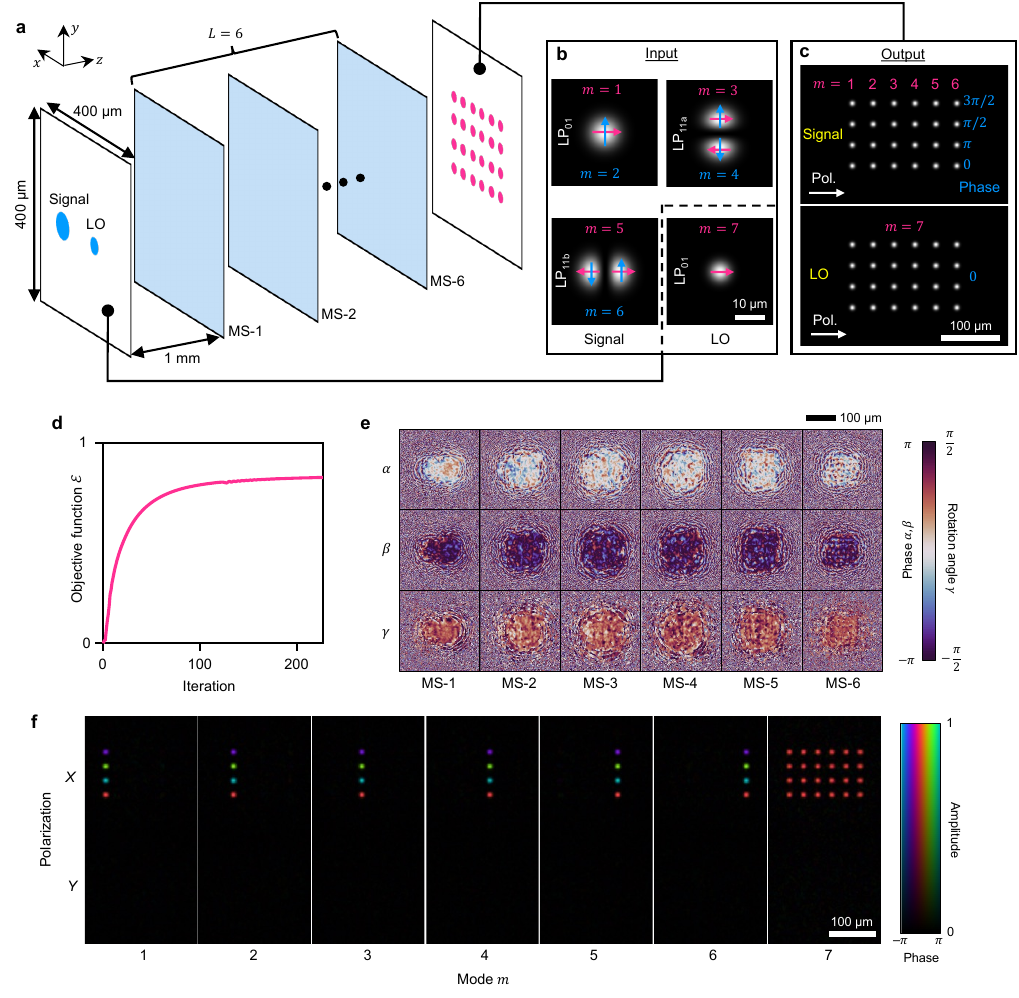}
\caption{
\textbf{MDM dual-polarization coherent receiver.} 
{\bf a}, Configuration of the device with six MS layers that realizes an optical hybrid for a mode-division-multiplexed (MDM) dual-polarization coherent receiver. We assume that each MS consists of meta-atoms arranged on a square lattice with a lattice constant of 1~{\textmu}m (i.e., $N=400^2$). Thus, the entire device contains 2,880,000 ($= 6\times400^2\times3$) parameters.
{\bf b}, Input mode profiles (LP$_{01}$, LP$_\mathrm{11a}$, and LP$_\mathrm{11b}$ for signals and LP$_\mathrm{01}$ for the local oscillator (LO)). 
{\bf c}, Target field profiles at the output plane for the signals and LO. The phase of each spot is set to achieve the functionality of a 90$^\circ$ optical hybrid for each mode.
The separation between adjacent spots and the beam diameter of each focused spot are set to 30~{\textmu}m and 10~{\textmu}m, respectively. 
{\bf d}, Objective function $\mathcal{E}$ during the optimization process. 
{\bf e}, Profiles of the optimized MS parameters $\alpha^{(l)}(\vb{r})$, $\beta^{(l)}(\vb{r})$, and $\gamma^{(l)}(\vb{r})$ ($l=1,2,\dots,6$). 
{\bf f}, Vectorial field distributions obtained at the output plane, $a_{m,X/Y}^\mathrm{out}(\vb{r})$ ($m=1,2,\dots,7$), for all input modes. For clarity, the optical phase offset in each field is adjusted.
}
\label{CoRx}
\end{figure*}

Here, we demonstrate a novel optical receiver frontend using our proposed vectorial mode converter that achieves all of the above functionalities, namely, a spatial mode demultiplexer, a PBS, and optical hybrids for all spatial/polarization modes, in one device. 
%The demonstrated device can be implemented in a folded MS structure as described in Fig.~\ref{concept}b, offering a cost-effective solution to realize compact MDM coherent receivers.
Figure~\ref{CoRx}a shows the configuration of the proposed MDM dual-polarization coherent receiver. 
%We assume a few-mode fiber (FMF) with a mode field diameter (MFD) of 15 {\textmu}m that supports dual-polarization signals in the three spatial modes (LP$_\mathrm{01}$, LP$_\mathrm{11a}$, and LP$_\mathrm{11b}$) at 1550-nm wavelength. 
We assume six layers of MSs ($L=6$) with 400-{\textmu}m-square area and a negligible thickness, each separated by 1~mm. In practice, the entire device can be implemented in a folded MS structure, offering a cost-effective solution to realize compact MDM coherent receivers.

% \begin{figure}[tb]
% \centering\includegraphics{figure_wExp/CoRx_setup.pdf}
% \caption{
% \textbf{MDM dual-polarization coherent receiver.} 
% {\bf a}, Configuration of the device with six MS layers. 
% {\bf b}, Input mode profiles (LP$_{01}$, LP$_\mathrm{11a}$, and LP$_\mathrm{11b}$ for signals and LP$_\mathrm{01}$ for LO). 
% {\bf c}, Target field profiles at the output plane for the signals and local oscillator (LO). The phase of each spot is set to achieve the functionality of a 90$^\circ$ optical hybrid for each mode.}
% \label{CoRx_setup}
% \end{figure}

Similar to the case considered in the previous section, we assume an FMF with an MFD of 15~{\textmu}m that supports dual-polarization signals in the three spatial modes (LP$_\mathrm{01}$, LP$_\mathrm{11a}$, and LP$_\mathrm{11b}$) at 1550-nm wavelength. 
As the LO light, we assume $X$-polarized LP$_\mathrm{01}$ mode from an SMF, having an MFD of 10~{\textmu}m. These fibers are placed at the input plane, with a separation of 127~{\textmu}m. For convenience, we define $m = 1,...,6$ to represent the six space/polarization modes of the signals and $m=7$ to represent the LO mode as shown in Fig.~\ref{CoRx}b.

Through adjoint optimization of meta-atoms, all lightwaves are converted to $X$ polarization and focused on 24 distinct points at the output plane, as shown in Fig.~\ref{CoRx}c. Here, each signal mode ($m = 1,...,6$) is focused on four vertically aligned points along its corresponding column. The optical phases at these four points are shifted by $\pi/2$. In contrast, the LO beam ($m = 7$) is split into 24 points with equal optical power and phases.  
As they interfere, therefore, we obtain the functionality of a 90$^\circ$ optical hybrid for each signal mode so that six coherent signals can be demodulated simultaneously by placing PDs at these 24 positions. 
%The separation between adjacent spots and the beam diameter of each focused spot are set to 30 {\textmu}m and 10 {\textmu}m, respectively. 
%We assume that each MS consists of meta-atoms arranged on a square lattice with a lattice constant of 1 {\textmu}m (i.e., $N=400^2$). 
% Thus, the entire device contains 2,880,000 ($= 6\times400^2\times3$) parameters. 
% In this work, the initial values of $\alpha^{(l)}(\vb{r}_n)$, $\beta^{(l)}(\vb{r}_n)$ and $\gamma^{(l)}(\vb{r}_n)$ are set to 0, $\pi/2$, and $\pi/4$, respectively, to prevent zero gradients. 
% We employ the band-limited angular spectrum method (ASM) for the free-space propagation \cite{Matsushima2009-ba}.
% and the adaptive moment estimation (ADAM) algorithm \cite{Kingma2014-hl} as an optimizer for efficient updating of parameters. 

% \begin{figure*}[t!]
% \centering\includegraphics{figure_wExp/CoRx_result.pdf}
% \caption{
% \textbf{Optimization results of MDM dual-polarization coherent receiver.} 
% {\bf a}, Objective function $\mathcal{E}$ during the optimization process. 
% {\bf b}, Profiles of the optimized MS parameters $\alpha^{(l)}(\vb{r})$, $\beta^{(l)}(\vb{r})$, and $\gamma^{(l)}(\vb{r})$ ($l=1,2,\dots,6$). 
% {\bf c}, Vectorial field distributions obtained at the output plane, $a_{m,X/Y}^\mathrm{out}(\vb{r})$ ($m=1,2,\dots,7$), for all input modes. For clarity, the optical phase offset in each field is adjusted. Scale bars in b and c are 100 {\textmu}m.
% }
% \label{CoRx_result}
% \end{figure*}

Figure~\ref{CoRx}d shows the obtained objective function $\mathcal{E}$ as a function of iteration, which shows good convergence after around 200 iterations.
% We perform parameter updating 200 times, enough to converge the objective function, as shown in Fig.~\ref{CoRx}(a).
Figures~\ref{CoRx}e and \ref{CoRx}f show the MS parameters of the optimized design and the vectorial field distributions at the output for each input mode, respectively (the evolution of vectorial field at each MS plane is provided in Supplementary Fig.~5). We can see that all modes are focused onto the well-defined positions with desired phases.

For quantitative evaluation, Table \ref{table1} summarises the performances of our designed MDM coherent receiver. Rigorous definitions and derivations of all metrics are given in the Methods section. 
We can see that the insertion loss is less than 0.9~dB for all modes with a low MDL of 0.1~dB. The crosstalk to other undesired PDs is suppressed below $-25$~dB, showing excellent spatial/polarization mode demultiplexing functions. Furthermore, the phase error and power imbalance within the four spots of each mode, which characterize the performance of the optical hybrid, are suppressed below 1.4$^\circ$ and 0.6~dB, respectively. 
%These values are comparable or superior to commercial 90$^\circ$ optical hybrids, which do not have the function of spatial mode demultiplexer \cite{Kylia, Optoplex, Finisar}. 
In addition, our device is confirmed to exhibit fairly robust operation across the entire $C$-band (1530-1565~nm) without significant degradation in performance.
Furthermore, increasing the number of layers generally leads to enhanced performance, owing to the larger degrees of freedom. 
(The wavelength dependence and effect of increasing the number of MS layers are examined in detail in Supplementary Notes~3 and 4, respectively.)

\begin{table*}[tb]
    \centering
    \caption{Performance of our MDM dual-polarization coherent receiver using the optimized MS parameters}
    \small
    \begin{ruledtabular}
    \begin{tabular}{lccccccccccc}
    % \hline
    LP mode && \multicolumn{2}{c}{LP$_\mathrm{01}$} &&  \multicolumn{2}{c}{LP$_\mathrm{11a}$} && \multicolumn{2}{c}{LP$_\mathrm{11b}$}  && LP$_\mathrm{01}$ (LO) \\ \cline{3-4}\cline{6-7}\cline{9-10}\cline{12-12}
    Polarization && $X$ & $Y$ && $X$ & $Y$ && $X$ & $Y$ && $X$ \\
    $m$ && 1 & 2 && 3 & 4 && 5 & 6 && 7\\ \hline
    Insertion loss (dB) && 0.81 & 0.82 &&	0.80 & 0.83 && 0.87 & 0.90	&& 0.78 \\
    %Efficiency (dB) && –0.81 & –0.82 &&	–0.80 & –0.83 && –0.87 & –0.90	&& –0.78 \\
    Crosstalk (dB)&&$-25.4$&$-26.0$&&$-30.1$&$-29.4$&&$-26.4$&$-28.0$&&——\\
    Phase error (deg) && $-0.29$ & $-0.26$ && $-0.28$ & 0.53 && 0.027 & $-0.27$ && 1.35 \\
%    Crosstalk (dB)&&–24.8&–25.4&&–28.9&–28.3&&–25.7&–27.0&&——\\
    Power imbalance (dB)&&0.36&0.25&&0.19&0.20&&0.075&0.066&&0.53 \\
    % \hline
    \end{tabular}
    \end{ruledtabular}
    \label{table1}
\end{table*}

\subsection*{Spatial-mode-multiplexed vectorial holography}
MS-based holography has been actively studied as a unique method to generate different images depending on the input and output polarization states \cite{Song2022-wb, Arbabi2015-rv, Balthasar_Mueller2017-ex, Bao2020-rj, Bao2021-yc, Bao2022-mz, Xiong2023-mk, Wen2015-ev, Zheng2022-cb}. By tuning the birefringence of each meta-atom, we can independently control the optical phase (and amplitude) distributions of $X$- and $Y$-polarization components and thereby synthesize separate images for respective polarizations at a desired plane. While the prior demonstrations mostly presumed an input beam with a single spatial mode, multiple spatial beams, such as Fourier modes \cite{Kamali2017-jt, Jang2021-ss, Deng2023-mj} and optical angular momentum (OAM) modes \cite{Jin2019-tl, Ren2019-cw, Ren2020-bs, Zhou2020-vl}, could also be used to produce spatially multiplexed images.
Such spatial-mode-multiplexed holography, however, has been restricted to specific input modes that possess some sort of spatial symmetry to enable simultaneous conversion using a mono-layer MS. 
Universal conversion of multiple beams, including their polarization profiles, for an arbitrary case of input vectorial modes would require transmission through multiple layers of MSs.
Here, we design our multi-layer MS device to demonstrate such an ultimate case of spatial-mode-multiplexed vectorial holography with high efficiency for the first time.

\begin{figure*}[tb!]
\centering\includegraphics{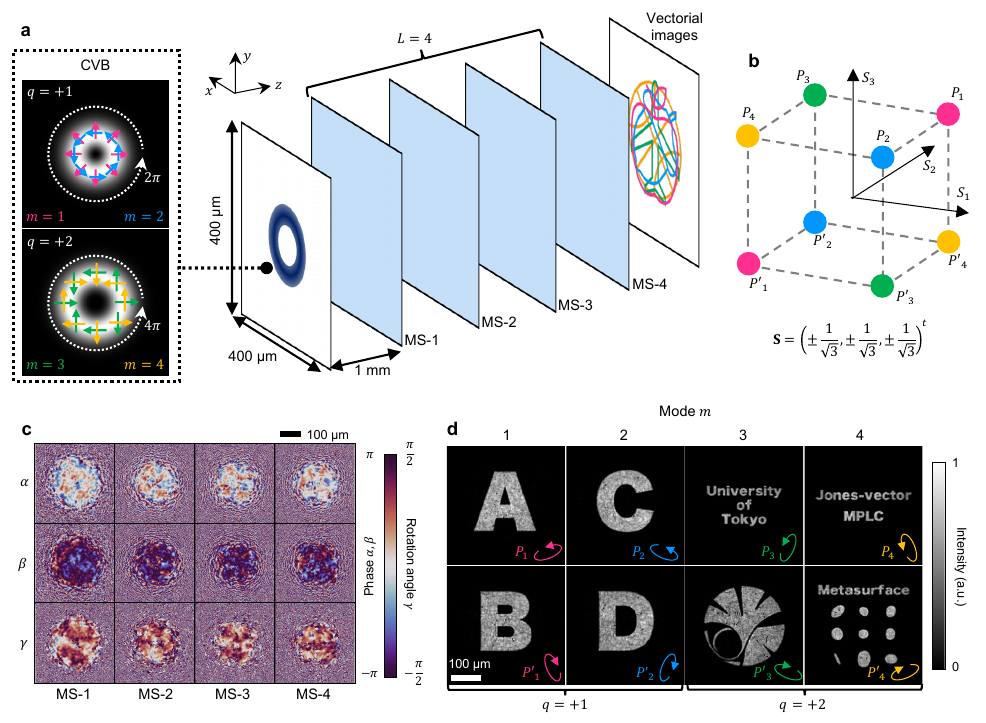}
\caption{
\textbf{Spatial-mode-multiplexed vectorial holography.} 
{\bf a}, Configuration of the spatial-mode-multiplexed vectorial holography with a four-layer MS. The left inset shows the input cylindrical vector beam (CVB) profiles with topological orders $q$ of +1 (top; $m=1,2$) and +2 (bottom; $m=3,4$). Arrows indicate the polarization orientation within the beam that rotates by $2\pi q$ along the azimuth.
{\bf b}, Analyzed polarization states $P_m$ and $P'_m$ ($m=1,2,3,4$) in the Stokes space, which composes a regular hexahedron. 
{\bf c}, Profiles of the optimized MS parameters $\alpha^{(l)}(\vb{r})$, $\beta^{(l)}(\vb{r})$, and $\gamma^{(l)}(\vb{r})$ ($l=1,2,3,4$) on 1-{\textmu}m-spacing grids. 
{\bf d}, Vectorial holographic images obtained at the output plane for two polarization states, $|a_{m,P_m}^\mathrm{out}(\vb{r})|^2$ and $|a_{m,P'_m}^\mathrm{out}(\vb{r})|^2$ ($m=1,2,3,4$). %Scale bars in c and d are 100 {\textmu}m.
}
\label{holography}
\end{figure*}

Figure~\ref{holography}a shows the configuration of the entire system, where the output image varies depending on the input vectorial mode and the analyzed polarization state. In this work, we select two CVB modes \cite{Zhan2009-gd} with the topological orders ($q$) of +1 and +2 as the input spatial modes.
Each CVB mode at 1550-nm wavelength is polarization-multiplexed so that we have four input modes ($m=1,2,3,4$) in total, as shown in the left inset of Fig.~\ref{holography}a.
The input beam diameter is set to 100~{\textmu}m for all modes.
After propagating through four layers of MSs ($L=4$), two different holographic images per each CVB mode are created depending on the analyzed polarization states, $P_m$ and $P'_m$. Here, $P_m$ and $P'_m$ represent the orthogonal polarization states that are analyzed for the $m$-th mode. We should note that the orthogonal polarization pair for each mode (${P_m}$, ${P'_m}$) can be freely selected, which is not possible with mono-layer MSs in principle \cite{Menzel2010-qo}. To demonstrate the versatility of our scheme, therefore, we set all four of them to be different elliptical polarization states; the Stokes parameters $\vb{S}=(S_1, S_2, S_3)^t$ of $P_m$ and $P'_m$ ($m=1,2,3,4$) are selected to constitute a regular hexahedron as shown in Fig.~\ref{holography}b.

The $m$-th target vectorial field at the output plane is then expressed as
\begin{equation}
    \vb{a}_{m}^\mathrm{tar}(\vb{r})=a_{m,P_m}^\mathrm{tar}(\vb{r})\ \vb{e}_{P_m}+a_{m,P'_m}^\mathrm{tar}(\vb{r})\ \vb{e}_{P'_m},
    \label{eq:vec_hol}
\end{equation}
where ${\vb{e}}_{P_m}$ and ${\vb{e}}_{P'_m}$ denote the unit Jones vectors of $P_m$ and $P'_m$, respectively. $|a_{m,P_m}^\mathrm{tar}(\vb{r})|^2$ and $|a_{m,P'_m}^\mathrm{tar}(\vb{r})|^2$ represent the $m$-th target holographic images for two orthogonal analyzing polarization states. 
Since we have the freedom to choose arbitrary phase profiles of the images, the phase distributions of $\vb{a}_{m}^\mathrm{tar}(\vb{r})$ are sequentially updated to match those of the output vectorial fields $\vb{a}_{m}^\mathrm{out}(\vb{r})$ after each forward calculation.
More specifically, $\vb{a}_{m}^\mathrm{tar}(\vb{r})$ is replaced to $\vb{a}_{m}^\mathrm{tar} (\vb{r}) e^{i\phi(\vb{r})}$ in every iteration, where $\phi(\vb{r})\equiv \mathrm{arg}[\{\vb{a}_{m}^\mathrm{tar}(\vb{r})\}^\dagger\vb{a}_{m}^\mathrm{out}(\vb{r})]$ is the phase difference between the output and target fields.
This operation corresponds to the Gerchberg–Saxton (GS) algorithm \cite{Gerchber1972-dm} and automatically ensures orthogonality between different target modes. Other parameters and optimization methods are the same as in the previous section.

Figures~\ref{holography}c and \ref{holography}d show the optimized MS parameters and simulated vectorial holographic images $|a_{m,P_m}^\mathrm{out}|^2$ and $|a_{m,P'_m}^\mathrm{out}|^2$ for each input mode ($m=1,2,3,4$). We should note that our method does not require a mode-selective aperture array at the output plane, unlike previously demonstrated OAM-multiplexed holography \cite{Jin2019-tl, Ren2019-cw, Ren2020-bs}. We can confirm that eight independent holographic images, including the fine texts and complex ginkgo mark of the University of Tokyo, are successfully generated. The holography efficiency, defined as $\qty|\bra{a_m^{\mathrm{tar}}}\ket{a_m^{\mathrm{out}}}|^2$, is as high as 93.5\%, 93.4\%, 88.8\%, and 86.8\% for $m=1$, $2$, $3$, and $4$, respectively.

\section*{Discussion and conclusion}
We have proposed and demonstrated a universal vectorial mode converter based on the MPLC concept with a multi-layer MS. The Jones matrix formalism was incorporated inside the conventional MPLC theory to enable local control of the polarization profiles of multiple beams in addition to their wavefronts. We then constructed a versatile inverse design algorithm based on the adjoint method to realize desired MIMO conversions of fully vectorial modes for arbitrary cases. 
The presented method was verified experimentally by demonstrating a 6-mode (LP$_\mathrm{01}$, LP$_\mathrm{11a}$, and LP$_\mathrm{11b}$ modes in $X$ and $Y$ polarization states) multiplexer using a 4-layer MS.
We designed and fabricated a compact device based on the folded MS configuration, and confirmed fully vectorial mode conversion of six $X$-polarized input beams to desired LP and polarization modes at the output plane.
Furthermore, the applicability of our scheme to more advanced MIMO devices was numerically demonstrated for two other use cases.
First, 
%the presented method was then applied to design 
a novel optical receiver frontend for MDM dual-polarization coherent signals was demonstrated; with the optimized MSs, we achieved simultaneous balanced homodyne detection of six coherent signals with excellent performances, such as 0.9-dB insertion loss, 0.1-dB MDL, and 25-dB crosstalk suppression for all modes. Second, we successfully demonstrated unique spatial-mode multiplexed vectorial holography using four CVB input beams and two arbitrary analyzing polarization states to generate eight independent images with more than 86\% efficiencies. 
%Finally, we experimentally demonstrate our approach by using the folded MS device, which was designed for the spatial and polarization mode multiplexer. We experimentally confirmed that the desired LP and polarization modes were observed at the output plane for all six input modes.

We should stress that such devices, capable of converting a set of multiple vectorial modes to another set of vectorial modes with arbitrary polarization profiles, are theoretically not possible with a mono-layer MS. 
This work, therefore, provides the first explicit and general formalism to realize universal MIMO vectorial mode converters. 
Moreover, we expect that independent wavelength-mode manipulation could also be incorporated into our device by using dispersive meta-atoms \cite{Zhou2020-vl, Miyata2021-cg, Faraji-Dana2018-xc, Faraji-Dana2019-ox, Chen2020-lv}. 
Owing to the versatility of the presented method, it can be applied to a variety of cases, paving the way toward the utilization of full DoFs of optical beams for diverse applications, including optical communication, imaging, and computing.

%%%%%%%%%% If using BibTeX:
\bibliography{references}

\newpage

\section*{Methods}
\subsection*{Metasurface design for mode multiplexer}
% For the MS optimization, the MS and simulation size was set to 300 {\textmu}m $\times$ 360 {\textmu}m, which was sufficiently larger than forward-propagated beams at the first MS layer and backward-propagated beams at the last MS layer. The discrete positions of meta-atoms and electric fields in simulation $\vb{r}_n$ were arranged on the square lattice with a lattice constant of 1.2 {\textmu}m. 
% For ease of calculation, the electric fields that propagate outside the lateral simulation region and reflect at the interface of air and SiO$_2$ were ignored, as well as the reflection loss at mirrors.
% In this folded MS configuration, where the light is incident to the MS with an oblique angle, we employed the shifted angular spectrum method for propagation \cite{Matsushima2010-wd}. 
% In the optimization process, the initial values of $\alpha^{(l)}(\vb{r}_n)$, $\beta^{(l)}(\vb{r}_n)$ and $\gamma^{(l)}(\vb{r}_n)$ are set to 0, $\pi/2$, and $\pi/4$, respectively, to prevent zero gradients. 
% We employ the adaptive moment estimation (ADAM) algorithm \cite{Kingma2014-hl} as an optimizer for efficient updating of parameters. 
% Other simulation and optimization methods were the same as in the others.

To design the MS for mode multiplexing shown in Fig.~\ref{MUX_sim}a, we first numerically derived the spatial distributions of design parameters ($\alpha$, $\beta$, and $\gamma$) required at four MS sections.
The discretization spacing of the position $\vb{r}_n$ in the simulation was set to 1.2~{\textmu}m (i.e., $N=250\times300$), which corresponded to twice the actual meta-atom spacing to be fabricated (0.6~{\textmu}m) to save the optimization cost.
Thus, the entire device contained 900,000 ($= 4\times250\times300\times3$) parameters. 
From the thickness of the SiO$_2$ substrate (625~{\textmu}m) and the size of each MS section (300~{\textmu}m), we set the incident angle to be $\tan^{-1} \{300~${\textmu}m$/(2\times625~${\textmu}m$\} \sim 13.5^\circ$. 
For calculating the wave propagation at an oblique angle, we employed the shifted angular spectrum method (ASM) \cite{Matsushima2010-wd}.
After iterative optimization (see Supplementary Fig.~1 for the detailed flow), we obtained Fig.~\ref{MUX_sim}b.

We then derived the actual dimensions $(D_\alpha, D_\beta, \gamma')$ of each meta-atom as shown in Fig.~\ref{MUX_exp}a to achieve the optimal parameters given in Fig.~\ref{MUX_sim}b. 
Note that due to the oblique incidence to the MS, $\gamma$ defined in Eq.~(\ref{eq:J_atom}) is not exactly the same as the geometrical in-plane rotation angle $\gamma'$ of meta-atoms. We, therefore, distinguish them using the prime symbol.
%The Si nanopost has an elliptical structure with dimensions $D_\alpha$ and $D_\beta$ as shown in Fig.~\ref{concept}a to induce birefringence. 
The meta-atoms were placed on a square lattice with a lattice constant of 0.6~{\textmu}m, which was less than the wavelength to avoid higher-order diffraction. 
% The optimized MS parameter distributions depicted in Fig.~\ref{MUX_sim}b were first interpolated onto a square grid with a 0.6 {\textmu}m spacing. Then, the meta-atom dimensions ($D_\alpha$, $D_\beta$) and rotation angle at each position were determined through rigorous coupled-wave analysis (RCWA). 
% From the optimized MS parameters on the 1.2-{\textmu}m-spaced grid $\vb{r}_n$, shown in Fig.~\ref{MUX_sim}b, we derived the parameters on the 0.6-{\textmu}m-spaced grid $\vb{r}'_n$, corresponding to the actual spacing of fabricated meta-atoms.
Therefore, new Jones matrix profiles $\vb{J}^{(l)}(\vb{r}'_n)$ on the square grid with 0.6~{\textmu}m spacing $(\vb{r}'_n)$ were obtained by interpolating
the optimal Jones matrices $\vb{J}^{(l)}(\vb{r}_n)$ on the 1.2-{\textmu}m-spacing grid $(\vb{r}_n)$ given in Fig.~\ref{MUX_sim}b. Since $\vb{J}^{(l)}(\vb{r}'_n)$ is not unitary in general, we approximated it with the unitary matrix $\vb{U}^{(l)}(\vb{r}'_n)$, which was obtained through polar decomposition of $\vb{J}^{(l)}(\vb{r}'_n)$.
The Jones matrix $\vb{J}(D_\alpha, D_\beta, \gamma')$ of the reflected light from a periodic array of meta-atoms was simulated using the rigorous coupled-wave analysis (RCWA) \cite{Liu2012-kd} for $s$- and $p$-polarization input at 1550-nm wavelength.
The refractive indices of SiO$_2$, Si, polyimide, and Au were set to 1.444, 3.48, 1.574, and 0.38+$i$10.75, respectively.
%, and the thicknesses of Si and polyimide between Si and Au were chosen to be 0.57 {\textmu}m and 1 {\textmu}m, respectively. 
Simulated Jones matrices in some parameters $(D_\alpha, D_\beta, \gamma')$ are shown in Supplementary Fig.~2.
Using these results, we derived geometrical parameters $(D_\alpha, D_\beta, \gamma')$ of the meta-atom at $\vb{r}’_n$ to minimize $\norm{\vb{U}^{(l)}(\vb{r}'_n)-\vb{J}(D_\alpha, D_\beta, \gamma')}^2$.

\subsection*{Device fabrication}
The fabrication flow is shown in Supplementary Fig.~3.
The folded MS device was fabricated on a silicon-on-quartz (SOQ) substrate with a Si thickness of 0.57~{\textmu}m. Positive EB resist (ZEP520A-7) was spin-coated, as well as the anti-charging conductive polymer (ESPACER 300Z). The MS pattern was written on the resist using an EB writer (ADVANTEST F7000S), the anti-charging layer was removed in de-ionized (DI) water, and the pattern was developed in a resist developer (ZED-N50). Then, the pattern was transferred to the Si layer using RIE with SF$_6$ and C$_4$F$_8$ gas, known as the Bosch process, followed by O$_2$ ashing process.
After the polyimide (Toray LT-S5181C) was spin-coated on the Si pattern, the device was annealed under a nitrogen atmosphere at 220~$^\circ$C for 1~hour for a cure. In preliminary experiments, we have confirmed the polyimide layer was embedded well, and the thickness of the polyimide layer on Si was around 1~{\textmu}m. Subsequently, a 200-nm-thick gold layer was deposited on the sample using a radio-frequency (RF) sputtering process. Then, the sample was flipped and cleaned by O$_2$ plasma ashing. The aperture patterns for input and output were written on the spin-coated photoresist using a laser writer with the back-side alignment to the MS patterns and then developed. Finally, mirror patterns were formed by a liftoff process after the deposition of silver and gold layers (200 and 50~nm) using an EB evaporator.

\subsection*{Measurement}
% The lightwave from the tunable laser source was split into two paths for the signal and reference through the 50/50 coupler. The signal light was incident to the device after the input polarization was set to X polarization by a polarization controller (PC). The position and angle of input SMF were modified by a precise 6-axis fiber stage. The output signal at the output plane was magnified ten times by using the 4-f systems with an objective lens (Mitsutoyo: M Plan Apo Nir) and a tube lens (Thorlabs: TTL200-S8) and imaged by an InGaAs camera (Artray: ARTCAM-991SWIR). To image the output intensity profiles for X and Y polarization, we inserted the polarizer and removed a beam splitter. On the other hand, to reconstruct the complex fields, we employed the off-axis digital holography. The reference signal was emitted into the space by a fiber collimator and interfered with the signal with a tilted angle by using a beam splitter. The polarization of the reference light was switched to X or Y polarization by using a PC depending on the polarization of the electric field to be reconstructed.
The laser light at 1550~nm was incident to the device from a fiber facet at an angle of 19.7$^\circ$ after the input polarization was set to $X$ polarization by a polarization controller (PC). The input mode ($m$) was controlled by moving the SMF using a 6-axis fiber stage to the corresponding position. The emitted beam at the output plane was magnified at ten times by using a 4-f system with an objective lens (Mitsutoyo: M Plan Apo NIR) and a tube lens (Thorlabs: TTL200-S8) and imaged by an InGaAs camera (Artray: ARTCAM-991SWIR). To image the output intensity profiles for $X$ and $Y$ polarization, we inserted a polarizer.

\subsection*{Performance metrics for MDM dual-polarization coherent receiver}
\label{sec:def-CR}
\subsubsection*{Coupling coefficients}
We define the coupling coefficient to an output spot at the $u$-th row ($u=1,2,3$, and $4$, corresponding to relative phase shifts of 0, $\pi$, $\pi/2$, and $3\pi/2$) and the $v$-th column ($v=1,2,\dots,6$) with $X$-polarization as
\begin{equation}
    c_{u,v,m} \equiv 
    \bra{g_{u,v}}\ket{a^\mathrm{out}_{m}}
    = \sum_n g^*_{u,v}(\vb{r}_n) a_{m,X}^\mathrm{out}(\vb{r}_n).
\end{equation}
where $\ket{g_{u,v}}=\sum_n g_{u,v}(\vb{r}_n)\ket{n,X}$ is a normalized $X$-polarized Gaussian field centered at the $(u,v)$-th spot and $a_{m,X}^\mathrm{out}(\vb{r}_n)=\bra{n,X}\ket{a^\mathrm{out}_{m}}$ is the $X$-polarized output field at $\vb{r}_n$ for the $m$-th input mode.

\subsubsection*{Insertion loss}
The total insertion loss (IL) for the $m$-th mode signal is derived from the sum of the coupling efficiencies to the four corresponding spots. We thus define IL as
\begin{equation}
    \mathrm{IL}_m \equiv 1 / \sum_{u=1}^4 \qty|c_{u,v=m,m}|^2.
%    \eta_m \equiv \sum_{u=1}^4 \qty|c_{u,v=m,p=X,m}|^2.
\end{equation}
For the LO ($m=7$), IL is defined as
\begin{equation}
\mathrm{IL}_\mathrm{LO} \equiv 1 /\sum_{u=1}^4 \sum_{v=1}^{6} \qty|c_{u,v,m=7}|^2.
    % \mathrm{IL}_\mathrm{LO} \equiv 1 /\sum_{u=1}^4 \sum_{v=1}^{M-1} \qty|c_{u,v,m=M}|^2.
%    \eta_\mathrm{LO} \equiv \sum_{u=1}^4 \sum_{v=1}^{M-1} \qty|c_{u,v,p=X,m=M}|^2.
\end{equation}

\subsubsection*{Phase error}
The phase error for the $m$-th mode signal is defined as the deviation from the ideal case of a 90$^\circ$ optical hybrid, which has $\pi/2$ phase difference between the in-phase and quadrature components \cite{Wen2020-ih}.
%\cite{Wen2019-gk}.
We thus have
\begin{equation}
    \qty|\Delta \phi_m| \equiv \qty| \angle (c_{4,m,m}-c_{3,m,m}) - \angle (c_{2,m,m}-c_{1,m,m}) -\pi/2 |.
\end{equation}
For the LO beam, the phase error in the output spots for the $m$-th mode is written as
\begin{equation}
    \qty|\Delta \phi_{\mathrm{LO},m}| \equiv \qty| \angle (c_{4,m,m}+c_{3,m,m}) - \angle (c_{2,m,m}+c_{1,m,m}) |.
\end{equation}
Then, we define the phase error for the LO as
\begin{equation}
    |\Delta \phi_{\mathrm{LO}}| \equiv \max_{m\in \{1,2,\dots,6\}} \qty|\Delta \phi_{\mathrm{LO},m}| .
    % |\Delta \phi_{\mathrm{LO}}| \equiv \max_{m\in \{1,2,\dots,M-1\}} \qty|\Delta \phi_{\mathrm{LO},m}| .
\end{equation}

\subsubsection*{Power imbalance}
The power imbalance for the $m$-th mode signal is defined as the ratio between the maximum and minimum values of the coupling efficiencies at the four spots in the corresponding column:
\begin{equation}
    \xi_m \equiv \frac{\underset{u}{\rm max}\ \qty|c_{u,v=m,m}|^2}{\underset{u}{\rm min}\ \qty|c_{u,v=m,m}|^2}.
\end{equation}
For the LO, the imbalance in the output spots for the $m$-th mode is written as
\begin{equation}
\xi_{\mathrm{LO},m} \equiv \frac{\underset{u}{\rm max}\ \qty|c_{u,v=m,m=7}|^2}{\underset{u}{\rm min}\ \qty|c_{u,v=m,m=7}|^2}.
    % \xi_{\mathrm{LO},m} \equiv \frac{\underset{u}{\rm max}\ \qty|c_{u,v=m,m=M}|^2}{\underset{u}{\rm min}\ \qty|c_{u,v=m,m=M}|^2}.
\end{equation}
Then, we define the imbalance for the LO as
\begin{equation}
    \xi_{\mathrm{LO}} \equiv \max_{m\in \{1,2,\dots,6\}} \xi_{\mathrm{LO},m} .
\end{equation}
\subsubsection*{Crosstalk}
The crosstalk for the $m$-th mode signal is defined as the sum of the coupling efficiencies at 20 undesired spots other than the four target spots,
\begin{equation}
    \chi_m \equiv \sum_{u=1}^4 \sum_{v\neq m}  \qty|c_{u,v,m}|^2.
\end{equation}

\section*{Data availability}  All the simulation codes and data presented in the paper are available upon a reasonable request.
% at https://doi.org/10.5281/zenodo.10158325 
%upon the publication.
%We have used the Matlab code in Ref. \cite{} as a great reference.

\section*{Acknowledgements}
This work was obtained in part from the commissioned research (No. JPJ012368C08801) by National Institute of Information and Communications Technology (NICT), Japan, and was partially supported by Japan Society of Promotion of Science (JSPS) KAKENHI, Grant Number 24KJ0557, and World-leading Innovative Graduate Study Program - Quantum Science and Technology Program (WINGS-QSTEP).
The SOQ wafer was provided by Shin-Etsu Chemical Co., Ltd. 
The device was fabricated in part at Takeda Cleanroom with help of Nanofabrication Platform Center of School of Engineering, the University of Tokyo, Japan, supported by "Advanced Research Infrastructure for Materials and Nanotechnology in Japan"
of the Ministry of Education, Culture, Sports, Science and Technology (MEXT) Grant Number JPMXP1224UT1115.
G.S. thanks Ryota Tanomura for the fruitful discussion. 

% G.S. acknowledges the financial support from World-leading Innovative Graduate Study Program - Quantum Science and Technology Fellowship Program (WINGS-QSTEP).

\section*{Author contributions}
G.S. conceived the idea and performed the simulation, MS design, device fabrication, measurement, and data analysis.
K.K. assisted G.S. with the MS fabrication. 
Y.N. contributed to the overall discussion and provided experiment facilities.
T.T. led the project and provided high-level supervision.
G.S. and T.T. wrote the manuscript.

\section*{Competing interests}
The authors declare no competing interests.

% \bibliography{references}
% \bibliographystyleappx{plain}
% \bibliographyappx{references}

\end{document}

\title{{\rm \it Supplementary Materials for}\\ Complete vectorial optical mode converter using multi-layer metasurface}
\author{Go~Soma}
\email{go.soma@tlab.t.u-tokyo.ac.jp}
\affiliation{School of Engineering, The University of Tokyo, Bunkyo-ku, Tokyo 113-8656, Japan}
\author{Kento~Komatsu}
\affiliation{School of Engineering, The University of Tokyo, Bunkyo-ku, Tokyo 113-8656, Japan}
\author{Yoshiaki~Nakano}
\affiliation{School of Engineering, The University of Tokyo, Bunkyo-ku, Tokyo 113-8656, Japan}
\author{Takuo~Tanemura}
\email{tanemura@ee.t.u-tokyo.ac.jp}
\affiliation{School of Engineering, The University of Tokyo, Bunkyo-ku, Tokyo 113-8656, Japan}
\maketitle
% \tableofcontents

\section{Derivation of Equation (13) and $\pdv{\vb{J}^{(l)}}{\theta^{(l)}}$}
Here, we derive Eq.~(13) in the main text.
From the objective function $\mathcal{E}$ defined in Eq.~(9), its derivative $\pdv{\mathcal{E}}{\theta^{(l)}(\vb{r}_n)}$ is expressed as
\begin{equation}
    \pdv{\mathcal{E}}{\theta^{(l)}(\vb{r}_n)}=\frac{1}{M}\sum_m 2\Re \qty[
    \braket{a_m^{\mathrm{out}}}{a_m^{\mathrm{tar}}}
    \mel**{a_m^{\mathrm{tar}}}{\pdv{\theta^{(l)}(\vb{r}_n)}}{a_m^{\mathrm{out}}}
    ],
    \label{eq:dEdp}
\end{equation}
where $\Re[z]$ describes the real part of the complex number $z$.
We should note that $\pdv{\theta^{(l)}(\vb{r}_n)}$ is effective only on $\hat{J}^{(l)}$ and not on other operators or $\ket{a_m^{\mathrm{tar}}}$, which are independent on the meta-atom structures. As a result, we obtain
\begin{equation}
    \begin{split}
        \mel**{a_m^{\mathrm{tar}}}{\pdv{\theta^{(l)}(\vb{r}_n)}}{a_m^{\mathrm{out}}}
        &=\mel**{a_m^{\mathrm{tar}}}{\pdv{\theta^{(l)}(\vb{r}_n)}\hat{F}^{(L)}\hat{J}^{(L)}\cdots \hat{F}^{(l)}\hat{J}^{(l)}}{a_m^{(l)}}\\
        % &=\mel**{a_m^{\mathrm{tar}}}{\hat{F}^{(l)}\hat{J}^{(l)}\cdots \hat{F}^{(l)}\pdv{\hat{J}^{(l)}}{\theta^{(l)}(\vb{r}_n)}}{a_m^{(l)}}\\
        &=\qty(\hat{F}^{(l)\dagger}\cdots\hat{J}^{(L)\dagger}\hat{F}^{(L)\dagger}\ket{a_m^{\mathrm{tar}}})^\dagger \pdv{\hat{J}^{(l)}}{\theta^{(l)}(\vb{r}_n)} \ket{a_m^{(l)}}\\
        &=\mel**{b_m^{(l)}}{\pdv{\hat{J}^{(l)}}{\theta^{(l)}(\vb{r}_n)}}{a_m^{(l)}}
    \end{split}
    \label{eq:dEdp2}
\end{equation}
Here, $\dagger$ is the Hermitian conjugate, so that $\hat{F}^{(l)\dagger}$ physically describes the backward propagation in the free space from ($l+1$)-th plane to $l$-th plane. $\langle{b_m^{(l)}}|$ is the adjoint vectorial field defined in Eq.~(14), which represents the vectorial field on the output of the $l$-th MS when the target field $|{a_m^{\mathrm{tar}}}\rangle$ is propagated backward. 

From the definition of the meta-atom operator $\hat{J}^{(l)}$ in Eq.~(5), $\pdv{\hat{J}^{(l)}}{\theta^{(l)}(\vb{r}_n)}$ in Eq.~(\ref{eq:dEdp2}) is written as
\begin{equation}
    \pdv{\hat{J}^{(l)}}{\theta^{(l)}(\vb{r}_n)} = \sum_{p,p'}\pdv{j_{pp'}^{(l)}(\vb{r}_n)}{\theta^{(l)}(\vb{r}_n)}(\vb{r}_n)\ketbra{n,p}{n,p'}.
    \label{eq:dJhatdp}
\end{equation}
Finally, by inserting Eqs. (\ref{eq:dEdp2}) and (\ref{eq:dJhatdp}) to Eq.~(\ref{eq:dEdp}) and using Eq.~(6), we obtain
\begin{equation}
\begin{split}
    \pdv{\mathcal{E}}{\theta^{(l)}(\vb{r}_n)}&=\frac{2}{M}\sum_{m} \Re \qty[
    \braket{a_m^{\mathrm{out}}}{a_m^{\mathrm{tar}}}
    \mel**{b_m^{(l)}}{\pdv{\hat{J}^{(l)}}{\theta^{(l)}(\vb{r}_n)}}{a_m^{(l)}}
    ]\\
    % &=\frac{1}{M}\sum_{m} 2\Re \qty[
    % \braket{a_m^{\mathrm{out}}}{a_m^{\mathrm{tar}}}
    % \sum_{p,p'}
    % \pdv{j_{pp'}^{(l)}(\vb{r}_n)}{\theta^{(l)}(\vb{r}_n)}
    % \braket{b_m^{(l)}}{n,p}
    % \braket{n,p'}{a_m^{(l)}}
    % ]\\
    % &=\frac{1}{M}\sum_{m} 2\Re \qty[
    % \braket{a_m^{\mathrm{out}}}{a_m^{\mathrm{tar}}}
    % \sum_{p,p'}
    % \pdv{j_{pp'}^{(l)}(\vb{r}_n)}{\theta^{(l)}(\vb{r}_n)}
    % \qty(b_{m,p}^{(l)}(\vb{r}_n))^*
    % a_{m,p'}^{(l)}(\vb{r}_n)
    % ]\\
    &=\frac{2}{M}\sum_{m} \Re \qty[
    \braket{a_m^{\mathrm{out}}}{a_m^{\mathrm{tar}}}
    \{\vb{b}_{m}^{(l)}(\vb{r}_n)\}^\dagger \pdv{\vb{J}^{(l)}(\vb{r}_n)}{\theta^{(l)}(\vb{r}_n)} \vb{a}_{m}^{(l)}(\vb{r}_n)
    ].
\end{split}
    \label{eq:dEdp3}
\end{equation}
Here, we define $\vb{a}_{m}^{(l)}(\vb{r}_n)$ and $\vb{b}_{m}^{(l)}(\vb{r}_n)$ as
\begin{equation}
    \vb{a}_{m}^{(l)}(\vb{r}_n) \equiv \mqty(a_{m,X}^{(l)}(\vb{r}_n)\\a_{m,Y}^{(l)}(\vb{r}_n)), \
    \vb{b}_{m}^{(l)}(\vb{r}_n) \equiv \mqty(b_{m,X}^{(l)}(\vb{r}_n)\\b_{m,Y}^{(l)}(\vb{r}_n)).    
\end{equation}
From the Jones matrix of a birefringent meta-atom described in Eq.~(8), $\pdv{\vb{J}^{(l)}}{\theta^{(l)}}$ can be expressed explicitly as
\begin{align}
    \pdv{\vb{J}^{(l)}(\vb{r}_n)}{\alpha^{(l)}(\vb{r}_n)}&=ie^{i\alpha^{(l)}(\vb{r}_n)}\mqty(\cos^2\gamma^{(l)}(\vb{r}_n) & \frac{1}{2}\sin 2\gamma^{(l)}(\vb{r}_n) \\ \frac{1}{2}\sin 2\gamma^{(l)}(\vb{r}_n) & \sin^2\gamma^{(l)}(\vb{r}_n)),
    \label{eq:dJdp1}\\
    \pdv{\vb{J}^{(l)}(\vb{r}_n)}{\beta^{(l)}(\vb{r}_n)}&=ie^{i\beta^{(l)}(\vb{r}_n)}\mqty(\sin^2\gamma^{(l)}(\vb{r}_n) & \frac{1}{2}\sin 2\gamma^{(l)}(\vb{r}_n) \\ \frac{1}{2} \sin 2\gamma^{(l)}(\vb{r}_n) & \cos^2\gamma^{(l)}(\vb{r}_n)),
    \label{eq:dJdp2}\\
    \pdv{\vb{J}^{(l)}(\vb{r}_n)}{\gamma^{(l)}(\vb{r}_n)}&=\qty(e^{i\alpha^{(l)}(\vb{r}_n)}-e^{i\beta^{(l)}(\vb{r}_n)})\mqty(-\sin 2\gamma^{(l)}(\vb{r}_n) & \cos 2\gamma^{(l)}(\vb{r}_n) \\ \cos 2\gamma^{(l)}(\vb{r}_n) & \sin 2\gamma^{(l)}(\vb{r}_n)).
    \label{eq:dJdp3}
\end{align}

\section{Measurement of complex field profiles and coupling matrix using off-axis holography}
To characterize the complex field profile of the output beam from the mode multiplexer,
%demonstrated in Section III of the main text,
we employed the off-axis holography method.
The measurement setup is shown in Supplementary Fig.~\ref{MUX_exp_S}(a). 
A continuous wave (CW) from a tunable laser source (TLS) at 1550~nm was split to the signal and reference paths through the 50/50 coupler. 
The signal light was incident to the device after the input polarization was set to $X$ polarization by a polarization controller (PC). 
The position and angle of input SMF were controlled by a 6-axis fiber stage. 
The signal at the output plane was magnified at ten times using a 4-f system with an objective lens (Mitsutoyo: M Plan Apo NIR) and a tube lens (Thorlabs: TTL200-S8). 
% Then, the sand imaged by an InGaAs camera (Artray: ARTCAM-991SWIR). 
Then, it was combined with the reference beam at a tilted angle by a beam splitter to generate interference fringes at an InGaAs camera (Artray: ARTCAM-991SWIR). The polarization of the reference light was switched to $X$ or $Y$ by using a PC to select each polarization component.

From the captured fringe patterns, we retrieved the complex field profiles $a_{m,p}^{\rm out, meas}(\vb{r})$ as shown in Supplementary Fig.~\ref{MUX_exp_S}(b).
Using these results, we calculated the coupling matrix $\vb{C}'$, where $C'_{mm'}=|\langle{a_m^{\mathrm{tar}}}|{a_{m'}^{\mathrm{out,meas}}}\rangle|^2$.
%($\ket{a_{m'}^{\mathrm{out,meas}}}$: the measured output mode for the $m'$-th mode input). 
Supplementary Fig.~\ref{MUX_exp_S}(c) shows the calculated coupling matrix.

\section{Wavelength dependence}
\begin{figure}[ht!]
\centering\includegraphics{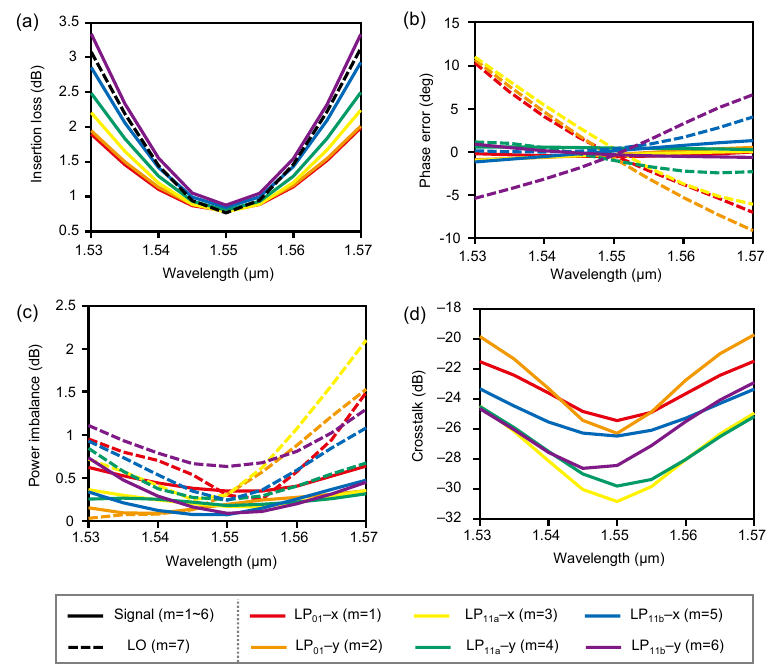}
\caption{Simulated wavelength dependence of the MDM dual-polarization coherent receiver: (a) Insertion losses, (b) phase errors, (c) power imbalances, and (d) crosstalk.}
\label{CoRx_wavelength}
\end{figure}

We investigate the wavelength dependence of the MDM polarization-diversity coherent receiver designed in %Section IV-A of 
the main text. 
Supplementary Fig.~\ref{CoRx_wavelength} shows the simulated performances as a function of wavelength. 
Here, we ignore the wavelength dependence of each meta-atom, which is a valid assumption for the wavelength range (1530-1570~nm) of our interest. 
Insertion loss, phase error, power imbalance, and crosstalk are suppressed below 3.5~dB, 11$^\circ$, 2.1~dB, and –19~dB at all wavelengths from 1530 to 1570~nm. These wavelength dependencies are expected to be improved further by employing the optimization at multiple wavelengths simultaneously. 

\section{Dependence of the number of metasurface layers}
\begin{figure}[ht!]
\centering\includegraphics{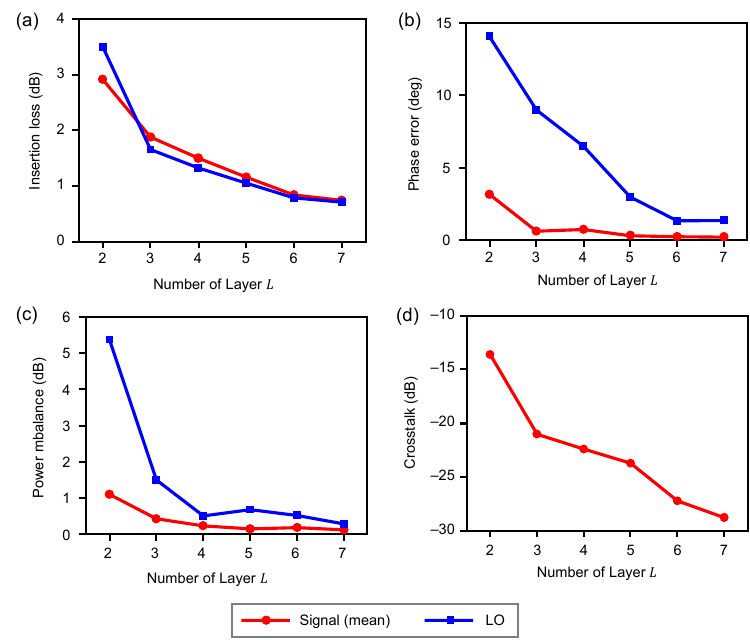}
\caption{Performance of the MDM dual-polarization coherent receiver as a function of the number of metasurface layers $L$: (a) efficiency, (b) phase error, (c) power imbalance, and (d) crosstalk. Averaged values of six modes are plotted for the signal.}
\label{CoRx_layer}
\end{figure}

We investigate the dependence of the number of metasurface layers on the performance of the MDM polarization-diversity coherent receiver. 
Supplementary Fig.~\ref{CoRx_layer} shows the simulated performances as a function of the number of metasurface layers. We can confirm that the performance generally improves by increasing the metasurface layers $L$.
In practice, however, optical insertion loss and sensitivity to fabrication errors increase as we increase the number of layers. There is, therefore, an optimal number of layers to maximize performance under actual conditions.

% For the calculation of the insertion losses, we first measured the optical power from the input SMF $P^{\rm in}$, and the optical power $P_{m=3}^{\rm out}$ just before the camera when the input SMF position was moved to the position corresponding to the third mode ($m=3$). 
% After measuring the optical transmission of the optics for measurement $T$, we calculated the transmission of the fabricated MS $T_{m=3}=T^{-1}P_{m=3}^{\rm out}/P^{\rm in}$. From this derived transmission $T_{m=3}$ and the output complex field reconstructed through off-axis holography, we derived the calibrated output complex field profile for the third mode input $a_{m=3, X}^{\rm out, meas}(\vb{r})$. Finally, we calculated the insertion loss of the third mode by $\qty|\bra{a_m^{\mathrm{tar}}}\ket{a_{m'}^{\mathrm{out,meas}}}|^2$. Insertion losses of the other modes were also derived from the normalized coupling matrix $\vb{C}'$ shown in Supplementary Fig.~\ref{MUX_exp_S}(c).

\clearpage

\begin{figure}[ht!]
\centering\includegraphics{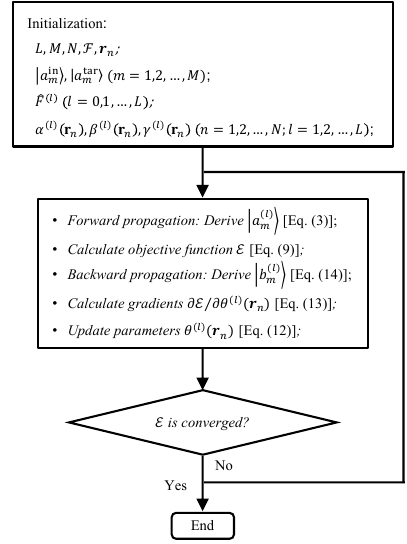}
\caption{Optimization flow of deriving the metasurface parameters. 
Initially, the MS parameters of $\alpha^{(l)}(\vb{r}_n)$, $\beta^{(l)}(\vb{r}_n)$ and $\gamma^{(l)}(\vb{r}_n)$ are uniformly set to 0, $\pi/2$, and $\pi/4$, respectively, to prevent zero gradients in this work.
In each iteration of optimization, we first calculate $|a_m^{(l)}\rangle$ for each mode by computing the forward propagation given by Eq.~(3) and derive $\mathcal{E}$ using Eq.~(9). 
Similarly, back-propagated fields $|b_m^{(l)}\rangle$ are obtained by computing Eq.~(14).
Here, we use the angular spectrum method (ASM) \cite{Matsushima2009-ba, Matsushima2010-wd} in the calculation of the free-space propagation.
Then, $\pdv{\mathcal{E}}{\theta^{(l)}(\vb{r}_n)}$ are calculated using Eq.~(13).
Finally, we update the parameters through Eq.~(12).
In this work, we use the adaptive moment estimation (ADAM) algorithm \cite{Kingma2014-hl} as an optimizer for efficient updating of parameters.
These procedures are repeated until the objective function converges
% In each iteration, we calculate the forward and the backward propagation, derive gradients, and update all parameters based on the gradient descent.
}
\label{optimization_flow}
\end{figure}

\begin{figure}[ht!]
\centering\includegraphics{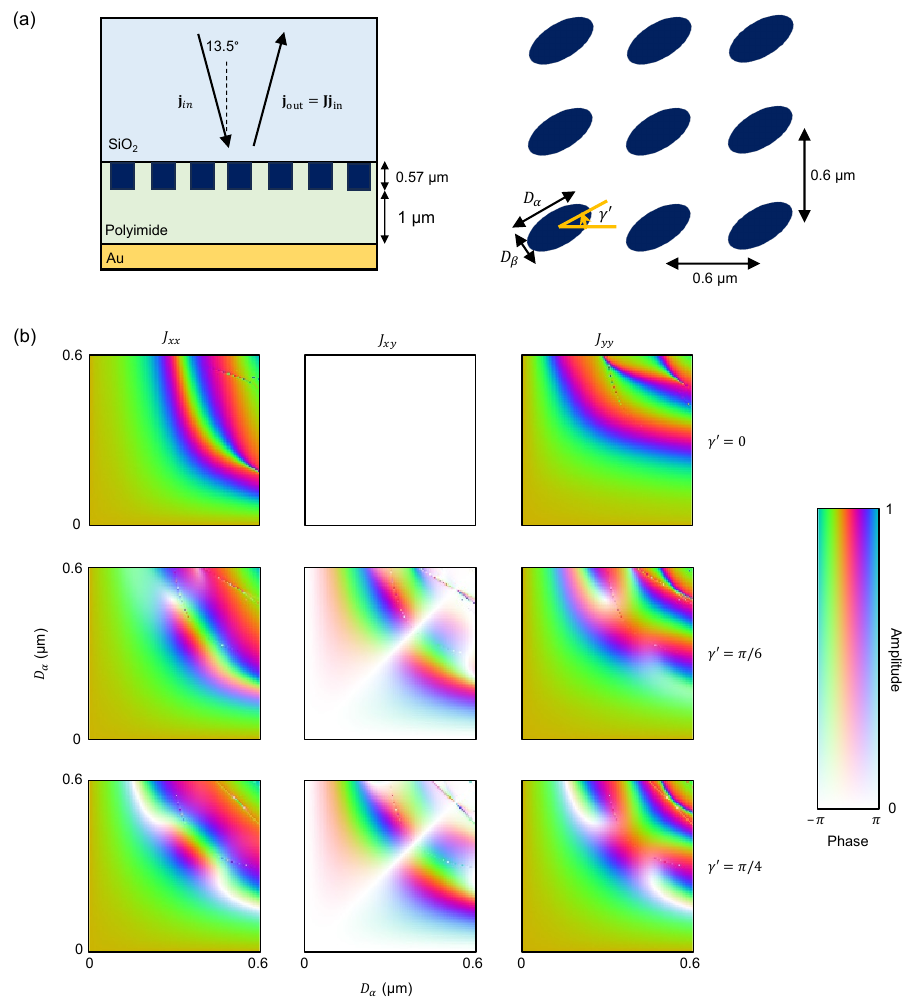}
\caption{(a) Schematic of a periodic reflective meta-atom array used to design the mode multiplexer. %demonstrated in Section III of the main text.
(b) Calculated Jones matrices of reflected light at 1550-nm wavelength as a function of $D_\alpha$, $D_\beta$, and $\gamma'$.}
\label{RCWA_sim}
\end{figure}

\begin{figure}[ht!]
\centering\includegraphics{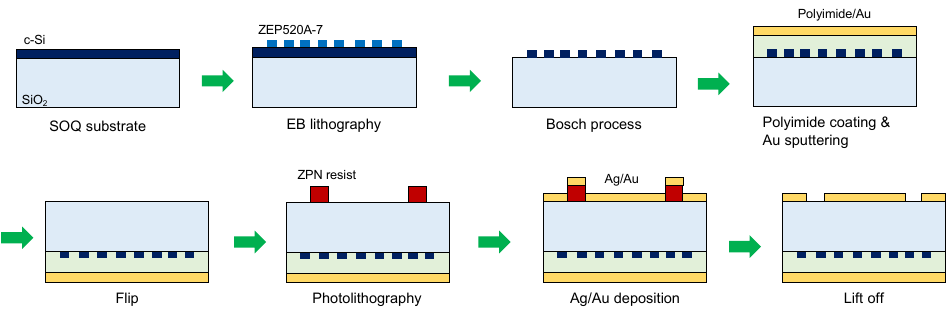}
\caption{
Schematic of the process flow used to fabricate the mode multiplexer chip. %demonstrated in Section III of the main text.
}
\label{process_flow}
\end{figure}

\begin{figure}[ht!]
\centering\includegraphics{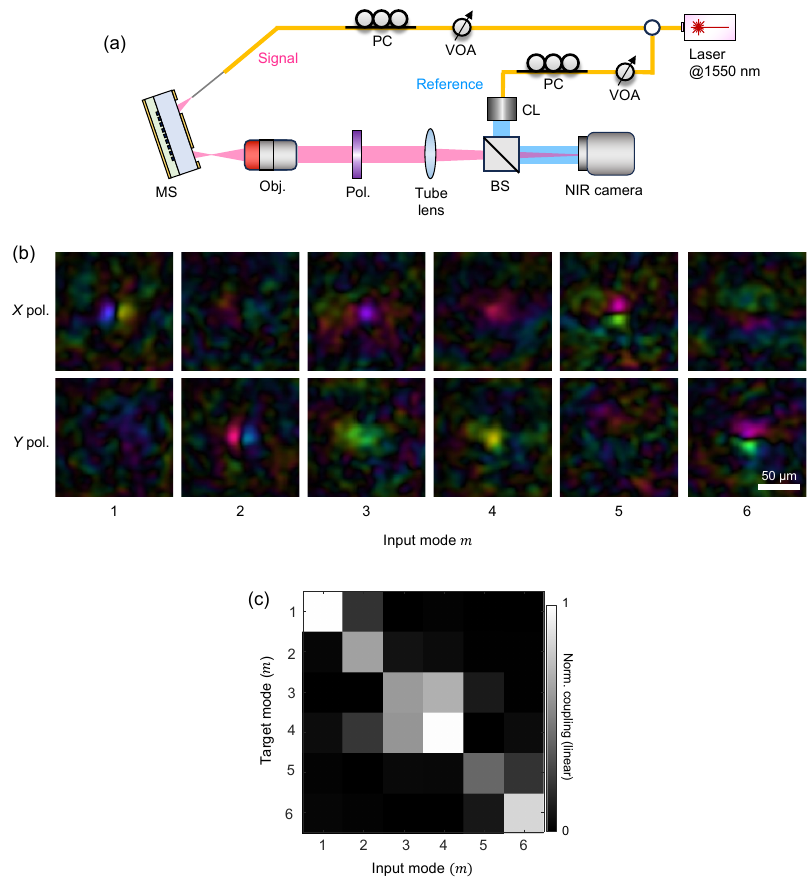}
\caption{(a) Measurement setup based on the off-axis holography. PC:~polarization controller. Obj.:~objective lens. Pol.:~polarizer. VOA:~variable optical attenuator. CL:~collimation lens. BS:~beam splitter.
(b) Reconstructed complex field profiles of the output beams for different input modes.
(c) Calculated coupling matrix.}
\label{MUX_exp_S}
\end{figure}

\begin{figure}[ht!]
\centering\includegraphics{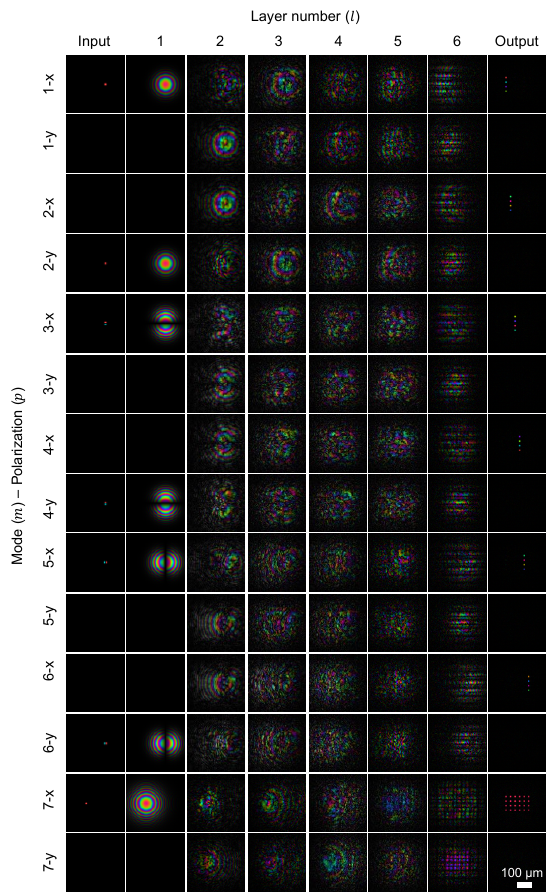}
\caption{Simulated field profiles $a_{m,p}^{(l)}(\vb{r})$ ($l={\rm in},1,2,\dots,6, {\rm out};\ m=1,2,\dots,7;\ p=X, Y$) of the MDM dual-polarization coherent receiver.% designed in Section IV-A of the main text.
}
\label{CoRx_field}
\end{figure}

\clearpage
\bibliography{references}